\documentclass[nofootinbib,aps,11pt,letterpaper,superscriptaddress]{revtex4}
\usepackage{graphicx}
\usepackage{hyperref}
\usepackage{cancel}
\usepackage{amssymb}
\usepackage{textcomp}
\usepackage{amsmath}
\usepackage{bm}
\usepackage{times}
\usepackage{epsfig}
\usepackage{color}
\usepackage{slashed}
\newcommand{\GeV}{{\rm GeV}}

\newcommand{\calW}{{\cal W}}

\usepackage{xcolor}

\begin{document}
	%\title{\LARGE  Dipole portal to heavy neutral leptons}
	\title{\LARGE  Probing active-sterile neutrino transition magnetic moments at LEP and CEPC}
	\bigskip
	\author{Yu Zhang}	
	\affiliation{School of Physics, Hefei University of Technology, Hefei 230601, P.R.China}

	\author{Wei Liu}	
	\email{wei.liu@njust.edu.cn, wei.liu.16@ucl.ac.uk}
	\affiliation{Department of Applied Physics and MIIT Key Laboratory of Semiconductor Microstructure and Quantum Sensing, Nanjing University of Science and Technology, Nanjing 210094, China}

	\begin{abstract}
		We consider the  sterile neutrino, that is also know as heavy neutral lepton (HNL), interacting with the Standard Model (SM) active neutrino and  photon via a transition magnetic moment, the so-called dipole portal, which can be induced from 
		the more general dipole couplings which respect the full gauge symmetries of the  SM.
		Depending on the interactions with 	${\rm SU}(2)_L$ and ${\rm U(1)}_Y$ field strength tensors $\calW_{\mu \nu}^a$ and $B^{\mu \nu}$, we consider four  typical scenarios and probe the constraints on the couplings with photon $d_\gamma$ at  LEP using the analyses  to search monophoton signature and the measurement of $Z$ decay. We find that in the considered scenarios assuming the coupling with $Z$ boson $d_Z\neq 0$, the measurement of $Z$ decaying into photon plus invisible particles can provide stricter constraints than the monophoton searches at the LEP1. The complementary constraints to existing experiments can be provided by the LEP.
		We also investigate the sensitivity on the dipole portal coupling $d_\gamma$ from the monophoton searches at future electron colliders, such as  CEPC, and find that  CEPC can explore the previously unconstrained parameter space by current experiments.		
	\end{abstract}
	
	\maketitle
	\newpage
	
	\section{Introduction}
	The discovery that neutrinos oscillate, and therefore have distinct mass and flavor eigenstates, has
	proven to be one of the most definitive pieces of evidence for physics beyond the Standard Model (BSM) in
	the last two decades \cite{SajjadAthar:2021prg, ParticleDataGroup:2022pth}. Given that the Standard Model (SM) does not predict the observed small and nonzero neutrino masses, 
	it is reasonable to introduce new physics which is typically organized in terms of the new particles and/or interactions.
	One feature present in many theories explain neutrino masses is the addition of one or more heavy neutral leptons (HNLs)
	$ N$, that can connect with neutrino masses via a Yukawa interaction,
	${\cal L}\supset NHL$, have attracted significant attentions in the last few years ~\cite{Balaji:2019fxd,Balaji:2020oig,Delgado:2022fea,Barducci:2022gdv,Ding:2019tqq,Shen:2022ffi,Deppisch:2018eth,Deppisch:2019kvs,Liu:2022kid,Liu:2021akf,Liu:2022ugx}. These neutral fermionic states $N$ are  singlet under the SM gauge groups, and often referred to as the so-called sterile neutrinos or right-handed neutrinos.

One of the consequences of extending the SM with additional sterile neutrinos, is
that the neutrino magnetic moment is generated with a tiny value proportional to the
neutrino mass \cite{Dvornikov:2003js,Shrock:1982sc,Pal:1981rm,Fujikawa:1980yx}. 
Recently, such scenarios predicting HNLs with the dipole coupling to SM active neutrinos, which are allowed to offer novel signatures and features in the production and decay of $N$ if the traditional neutrino portal coupling $NHL$ is  assumed to be absent or subdominant, have received renewed attention and have been studied in the context of colliders, beam-dump and neutrino experiments, astrophysics, cosmology, and direct
searches at dark matter experiments \cite{Aparici:2009fh, Giunti:2014ixa, Aparici:2013xga, Coloma:2017ppo, Abazajian:2017tcc,Shoemaker:2018vii, Magill:2018jla, Brdar:2020quo, Plestid:2020vqf,Jodlowski:2020vhr, Schwetz:2020xra,  Ismail:2021dyp,Miranda:2021kre,Dasgupta:2021fpn,Atkinson:2021rnp,Kamp:2022bpt,Gustafson:2022rsz,Huang:2022pce,Li:2022bqr,Acero:2022wqg,Feng:2022inv,Hati:2022tfo,Mathur:2021trm,Bolton:2021pey,Ovchynnikov:2022rqj,Zhang:2022spf,Guo:2023bpo,Alok:2022pdn}.
At the effective low-energy level, neutrino dipole portal to HNLs is described by the
 Lagrangian
\begin{equation}
	\mathcal{L} \supset d_k \bar{\nu}_{L}^k \sigma_{\mu \nu} F^{\mu \nu} N+\mathrm{H.c.},
\label{eq:L}
\end{equation}
where $k=e,\mu,\tau$ denotes the flavor index of lepton,
$\nu_L$ is a SM left-handed (active) neutrino field,
$\sigma_{\mu \nu}=\frac{i}{2}[\gamma_\mu,\gamma_\nu]$, 
$F^{\mu \nu}$ is the electromagnetic field strength tensor, and $d$ is the active-sterile neutrino transition magnetic moment, that controls the strength of the interaction with the units of (mass)$^{-1}$.

If the typical momentum exchange is much smaller than the electroweak scale, only considering the dipole portal coupling to sterile-active neutrinos and electromagnetic field strength tensor in the simplified model in Eq. (\ref{eq:L}) at the effective low-energy level is suitable. While scattering energy can be comparable with or above the electroweak scale, the SM gauge invariant dipole couplings should be taken into account \cite{Magill:2018jla, Butterworth:2019iff}. 
The main aim of this study is to investigate the  active-sterile neutrino transition magnetic moments which respect the full gauge symmetries of the SM, and to probe the corresponding sensitivity at the electron colliders with the center-of-mass (CM) energy $\sqrt{s}\gtrsim M_Z$, such as  LEP and future Circular Electron Positron Collider (CEPC)~\cite{CEPCStudyGroup:2018rmc, CEPCStudyGroup:2018ghi} .

The paper is organized as follows. In section II, we describe the model include the effective Lagrangian for the dipole portal coupling to HNLs. The production of sterile neutrino $N$ at electron colliders is investigated in section III. We then discuss the constraints
from LEP in section IV, and from future CEPC in section V.  Section VI contains our discussion and conclusion.

\section{The model}

It is worth noting that the effective Lagrangian in Eq. (\ref{eq:L}), describing the active-sterile neutrino transition magnetic moments, is not gauge invariant under ${\rm SU}(2)_L\times{\rm U(1)}_Y$ gauge group.
In order to describe the new physics above the EW scale, neutrino dipole couplings which respect the full gauge symmetries of the standard model are need to be considered and can be written as \cite{Magill:2018jla}
\begin{equation}
\mathcal{L} \supset \bar{L}^k(d_{\calW}^k \calW_{\mu \nu}^a \tau^a + d_B^k B^{\mu \nu}) \tilde{H}\sigma_{\mu \nu} N+\mathrm{H.c.},
\label{eq:LWB}
\end{equation}
where $\tilde{H}=i\sigma_2H^*$, $\tau^a=\sigma^a/2$ with $\sigma^a$ being Pauli matrices, $\calW_{\mu \nu}^a$ and $B^{\mu \nu}$ denote the
${\rm SU}(2)_L$ and ${\rm U(1)}_Y$ field strength tensors with $\calW_{\mu \nu}^a\equiv\partial_\mu \calW_\nu^a-\partial_\nu \calW_\mu^a+g\epsilon^{abc}W_\mu^b W_\nu^c$ and $B_{\mu\nu}\equiv\partial_\mu B_\nu-\partial_\nu B_\mu$.
It can be seen that because of  a Higgs insertion, the dipole interaction in Eq. (\ref{eq:LWB}) 
is really a dimension 6 operator.
{The Lagrangian in Eq.(\ref{eq:LWB}) can also described with the Wilson coefficients $C_B$ and $C_{\calW}$ by the replacement of  $d_B\sim \frac{C_B}{\Lambda^2}$ and $d_{\calW}\sim \frac{C_{\calW}}{\Lambda^2}$,  where $\Lambda$ is the cutoff energy scale.}

After electroweak symmetry breaking (EWSB) with the Higgs vacuum expectation value $v$, one obtains 
\begin{equation}
\mathcal{L} \supset d_W^k(\bar{\ell}^k W^-_{\mu \nu} \sigma_{\mu \nu} N) + \bar{\nu}^k_L(d_\gamma^k F_{\mu \nu}-d_Z^k Z_{\mu \nu}) \sigma_{\mu \nu} N +\mathrm{H.c.},
\label{eq:LWZk}
\end{equation}
which can induce  dipole operators to SM photon, the weak boson $Z$ and $W$, with the coupling $d_\gamma^k$, $d_Z^k$ and $d_W^k$. One sees that the normalization of the
photon field strength term in Eq. (\ref{eq:LWZk}) can induce that of Eq. (\ref{eq:L}).
{The  active to sterile transition magnetic moment $d_\gamma^k$ is anticipated to be of order $\sim \frac{e v}{\Lambda^2}$, where  $v$ is
	the Higgs field vacuum expectation value. }
For a given lepton flavor, the dipole couplings $d_\gamma$, $d_Z$ and $d_W$ in the broken phase are linearly dependent,
and determined by only two parameters $d_{\calW}$ and $d_B$ in the unbroken phase via
\footnote{ As
	we will mostly assume that only one of the active neutrino flavors participates in the magnetic moment interactions,   the superscript $k$ of the lepton flavor will be omitted in the following to simplify our notation, unless otherwise stated. }
\begin{eqnarray}
d_\gamma&=&\frac{v}{\sqrt{2}}\left(d_B\cos\theta_{w}+\frac{d_{\calW}}{2}\sin\theta_w\right), \nonumber\\ 
d_Z&=&\frac{v}{\sqrt{2}}\left(\frac{d_{\calW}}{2}\cos\theta_{w}-d_B\sin\theta_w\right), \nonumber\\ 
d_W&=&\frac{v}{\sqrt{2}}\frac{d_{\calW}}{2}\sqrt{2}.
\end{eqnarray}

One find that, we have three free parameters in this model
%The independent parameters of the model considered in this work are
\begin{equation}\label{eq:freepara1}
	\{m_N, d_{\calW}, d_B\},
\end{equation}
where $m_N$ is the mass of HNL.
Assuming $d_{\calW}=a\times d_B$, we have
\begin{eqnarray}
d_Z&=&\frac{d_\gamma(a\cos\theta_w-2\sin\theta_w)}{2\cos\theta_w+a\sin\theta_w}, \nonumber \\
d_W&=&\frac{\sqrt{2}a d_\gamma}{2\cos\theta_w+a\sin\theta_w}.
\end{eqnarray}
Then the independent parameters of (\ref{eq:freepara1}) can be replaced  by the parameters
\begin{equation}
\{m_N, d_\gamma, a\}.
\end{equation}
In this work, we will focus on four typical scenarios, which are listed in Table \ref{tab:scen}.
%In scenario I,  $a=0$ means that HNL doesn't couple with the isotriplet $\calW_\mu^a$ of the group ${\rm SU}(2)_L$ leading to the coupling with $W$ boson $d_W=0$. In scenario II, HNL doesn't couple with the isosinglet $B_\mu$ of the group ${\rm U(1)}_Y$, which can be understood as $d_B/d_\calW=0$ corresponding to $a\to\infty$.
%In scenario III, $a=2\tan\theta_w$ results in that the coupling with $Z$ boson $d_Z$ vanishes. For comparison, in scenario IV we set  $a=-2\tan\theta_w$, which is the negative value in scenario III. 
{In order to investigate the operators with $B_\mu$ and ${\mathcal W}_\mu$	individually, we consider $d_B=0\ (C_B=0)$ or   $d_{\mathcal W}=0\ (C_{\mathcal W}=0)$ in scenario I and II, respectively.
	One see that, scenario I with  $a=0$ leads to  the coupling with $W$ boson $d_W=0$ after EWSB. With $d_B=0$, scenario II corresponds to  $a=d_{\mathcal W}/d_B \to\infty$. We also consider scenario III in which  $a=2\tan\theta_w$ results in that the coupling with $Z$ boson $d_Z$ vanishing. For comparison, in scenario IV we set  $a=-2\tan\theta_w$, which is the negative value in scenario III. }

\begin{table}[]
	\centering
	\begin{tabular}{c|c|c} \hline\hline
		Scenario &  Assumptions   &  Relations\\ \hline\hline
		I &  $d_{\calW}=0$   & $d_Z=-d_\gamma \tan\theta_w$; $d_W=0$\\ \hline
		II &  $d_{B}=0$   & $d_Z=d_\gamma \cot\theta_w$; $d_W=\sqrt{2}d_\gamma/\sin\theta_w$\\ \hline
		III &  $d_{\calW}=2\tan\theta_w\times d_B$ & $d_Z=0$;  $d_W=\sqrt{2}d_\gamma\sin\theta_w$ \\ \hline
		%			IV &  $d_{\calW}=a\times d_B$ & $d_Z=\frac{d_\gamma(a\cos\theta_w-2\sin\theta_w)}{2\cos\theta_w+a\sin\theta_w}$;  $d_W=\frac{\sqrt{2}a d_\gamma}{2\cos\theta_w+a\sin\theta_w}$ \\\hline\hline
		IV &  $d_{\calW}=-2\tan\theta_w\times d_B$ &  $d_Z=-d_\gamma \tan(2\theta_w)$;  $d_W=-{\sqrt{2} d_\gamma \sin\theta_w}/{\cos(2\theta_w)}$ \\\hline\hline
	\end{tabular}
	\caption{Four typical scenarios considered in this work.}
	\label{tab:scen}
\end{table}

	{In general, the gauge invariant operators can generate neutrino masses at the dim-6 level with a Majorana mass term  $\mathfrak{m}_N$ via loop diagrams. In this work, we consider the sterile neutrino $N$ as Dirac fermion, then the sterile neutrino  is decoupled from the mechanism that generates active neutrino masses.  Large dipole couplings can be made compatible with small  neutrino masses if the $N$ is Dirac, or quasi-Dirac with a small Majorana-type mass splitting satisfying $\mathfrak{m}_N\ll m_N$ \cite{Magill:2018jla}.	
}

		The sterile neutrino $N$ as a Dirac fermion can decay into an on-shell vector boson and a SM lepton 
	through the dipole operators in Eq. (\ref{eq:LWZk}), 
	with the decay rates given by
	\begin{eqnarray}
		\Gamma_{N\to\nu\gamma}&=&\frac{d_\gamma^2m_N^3}{4\pi}, \nonumber \\
		\Gamma_{N\to\nu Z}&=&\frac{d_Z^2(m_N^2-M_Z^2)^2(2m_N^2+M_Z^2)}{8\pi m_N^3} \Theta(m_N>M_Z), \nonumber \\
		\Gamma_{N\to W\ell}&=&\frac{d_W^2}{8\pi m_N^3} \sqrt{(m_N^2-(M_W-m_\ell)^2)(m_N^2+(M_W-m_\ell)^2)} \nonumber\\
		&\times&\left(2m_\ell^2(2m_\ell^2-4m_N^2-M_W^2)+(m_N^2-M_W^2)(2m_N^2+M_W^2)\right) \Theta(m_N>M_W+m_\ell).
	\end{eqnarray}
Besides, there will be three-body decay channels of HNL via off-shell $W$ and $Z$ bosons, such as $N\to W^*\ell\to\ell+ff^\prime$ and $N\to\nu Z^*\to\nu f\bar{f}$, which are suppressed by a factor of fine structure constant $\alpha$. While when $d_W$ or $d_Z$ is  much larger than $d_\gamma$, these three-body decay channels can play an important role with $m_N < m_W$, which will reduce the branching ratio of $N\to\nu\gamma$.
The Universal FeynRules Output~(UFO)~\cite{Alloul:2013bka, Degrande:2011ua} of the neutrino dipole model is used, and fed to {\tt MadGraph5aMC$@$NLO} -v2.6.7~\cite{Alwall:2014hca} to calculate the width of three-body decay channels of HNL.

\begin{figure}[htbp]
		\begin{centering}			
			\includegraphics[width=0.45\columnwidth]{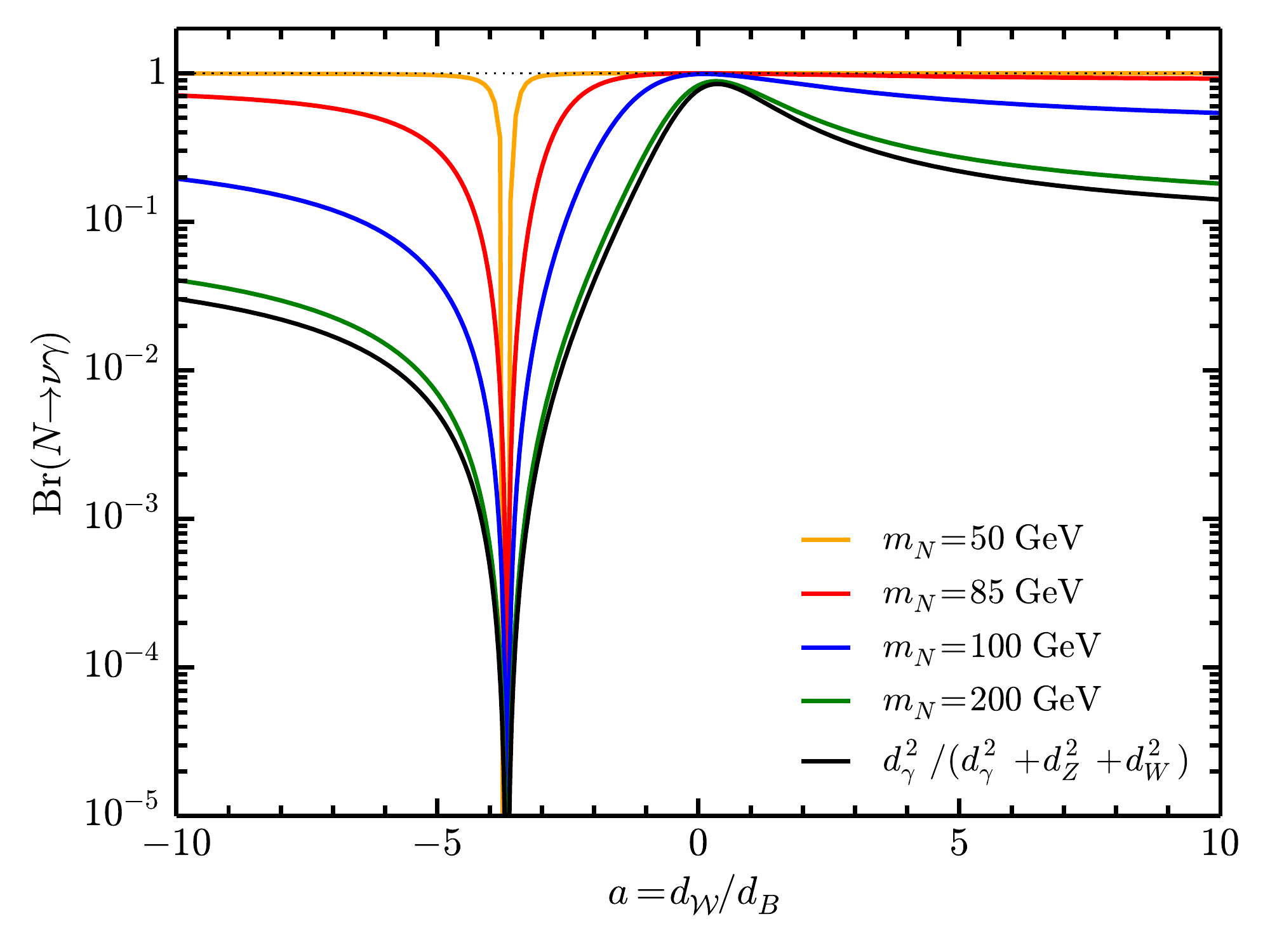}
			\includegraphics[width=0.45\columnwidth]{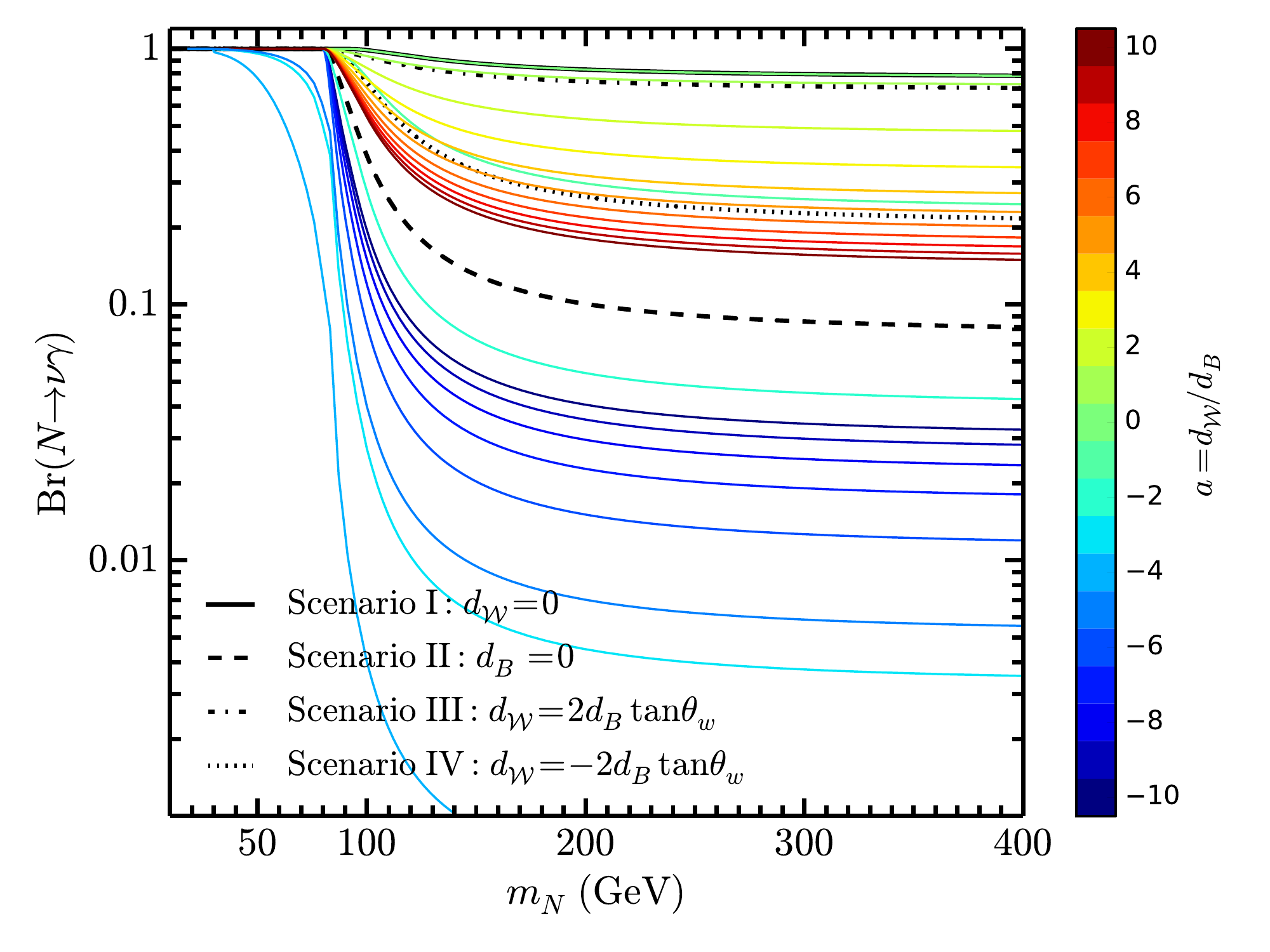}
			\caption{Branching ratio of the radiative HNL decay process $N\to\nu\gamma$ as  the ratio $a=d_{\calW}/d_B$ (left) and the function of the HNL mass $m_N$ (right).}
			\label{fig:br}
		\end{centering}
	\end{figure}

In Fig. \ref{fig:br}, we present the branching ratio for $N$ decaying to a photon plus active neutrino 
\begin{equation}
{\rm Br}(N\to\nu\gamma)\equiv\frac{\Gamma_{N\to\nu\gamma}}{\Gamma_{N\to\nu\gamma}+\Gamma_{N\to\nu Z}+\Gamma_{N\to W\ell}+\Gamma_{N\to \text{3-body}}} 
\end{equation}
as the function of $m_N$ and the ratio of $a=d_{\calW}/d_B$. 
From the  curves in $a-{\rm Br}(N\to\nu\gamma)$ plane, we can find a very conspicuous sharp valley, because there is a singularity in $d_Z$ and $d_W$ with $a=-2\cot\theta_{w}\sim-3.7$. 
Around this singularity point, the three-body decay channels of HNL via off-shell $W$ and $Z$ bosons  can be sizable and even dominant when  $m_N<M_W$.
 While the four scenarios listed in Table \ref{tab:scen} are all away from the singularity point $a\sim-3.7$, thus the three-body decay channels can be safely ignored in our following calculations. 
From the curves in $m_N-{\rm Br}(N\to\nu\gamma)$ plane, it can be seen that  the branching ratio always decrease  with the increment of $m_N$ when $m_N>M_W$.  Since the width of three-body decay channels can be neglected compared with the two-body decays when $m_N>M_W$,  ${\rm Br}(N\to\nu\gamma)$  tends to be $d_\gamma^2/(d_\gamma^2+d_Z^2+d_W^2)=(4+3a^2)/(2\cos\theta_w+a\sin\theta_w)^2$. 
For the heavy neutrino with $M_N\gg M_Z$, the branching ratio ${\rm Br}(N\to\nu\gamma)$ will reach its maximum near $a =0$, then decreases with the increment of $a$, and finally tends to be that obtained in Scenario II with $d_B=0$.

%%%%%%%%%%%%%%%%%%%%%%%%%%%%%%%%%%%%%%%%%%%%%%%%%%%%%%%%%%%%%%%
%%%%%%%%%%%%%%%%%%%%%%%%%%%%%%%%%%%%%%%%%%%%%%%%%%%%%%%%%%%%%%%
	
	\section{Electron collider signals}
	\label{sec:sb}
	In this section, we will investigate the Dirac sterile {neutrino} $N$ production via dipole portal at
	high energy $e^+e^-$ colliders, such as LEP,  and future CEPC.
	At electron colliders, 	HNL production will proceed from the process $e^+e^-\to N\bar\nu_k +{\rm H.c.}$  via either $Z$ or $\gamma$ mediator in $s$-channel depending on dipole portal couplings $d_Z^k$, $d_\gamma^k$ with $k=e,\mu,\tau$, 
	 or  via $W$ mediator in $t$-channel depending on electron neutrino dipole portal coupling  $d_W^e$ in Eq. (\ref{eq:LWZk}), respectively . 
	With the subsequent decay channel $N\to\nu\gamma$ in the detector, the signature of a single photon final state with missing energy can be look for at  electron colliders.
	The total production cross section for 
	$e^+e^-\to N \bar\nu$  after integrating over 
	all angles  from $\gamma$ and $Z$ mediators in $s$-channel and $W$ mediator in $t$-channel 
	can be respectively expressed as 
	\begin{eqnarray}
		\sigma^\gamma(e^+e^-\to N \bar\nu)&=&\frac{\alpha d_\gamma^2 (s-m_N^2)^2(s+2m_N^2)}{3 s^3},  \\ 
		\sigma^Z(e^+e^-\to N \bar\nu)&=&\frac{\alpha d_Z^2(s-m_N^2)^2(s+2m_N^2)}{24 c_w^2  s_w^2 s \left(\Gamma_Z^2 M_Z^2+\left(M_Z^2-s\right)^2\right)} \Big[  \left(8 s_w^2-4 s_w+1\right)\Big], \\ 
		\sigma^{\gamma Z}(e^+e^-\to N \bar\nu)&=&\frac{\alpha  d_\gamma d_Z (s-m_N^2)^2(s+2m_N^2)}{6 c_w  s_ws^2 \left(\Gamma_Z^2 M_Z^2+\left(M_Z^2-s\right)^2\right)}  \Big[
		(-4 s_w+1) \left(M_Z^2-s\right)\Big], \\ 	
		\sigma^W(e^+e^-\to N \bar\nu_e)&=&\frac{\alpha (d_W^e)^2}{2    s_w^2s} \Bigg[-2 s-\left(2 M_W^2+s\right) \log \left(\frac{M_W^2}{-m_N^2+M_W^2+s}\right) \nonumber\\ 
		&+&m_N^2 \left(\frac{M_W^2}{-m_N^2+M_W^2+s}+1\right)\Bigg],
		\label{eq:sigma}
	\end{eqnarray}
	where $\sigma^{\gamma Z}$ denotes the interference term between $\gamma$ and $Z$ mediators in $s$-channel, and 
	the interference between $s$-channel and $t$-channel for electron neutrino vanished. 
	One sees that at the low-energy electron colliders with $\sqrt{s}\ll M_{Z}$, the contribution from $Z$ or $W$ mediator can be
 neglected comparing with the one from $\gamma$ mediator in the condition of ${d_{Z,W}}/{d_\gamma}\sim \mathcal{O}(1)$, which has been  discussed in Ref.\cite{Zhang:2022spf} at the low-energy electron colliders, such as BESIII, Belle II and future STCF. 
 
\begin{figure}[htbp]
 	\begin{centering}
 		\includegraphics[width=0.45\columnwidth]{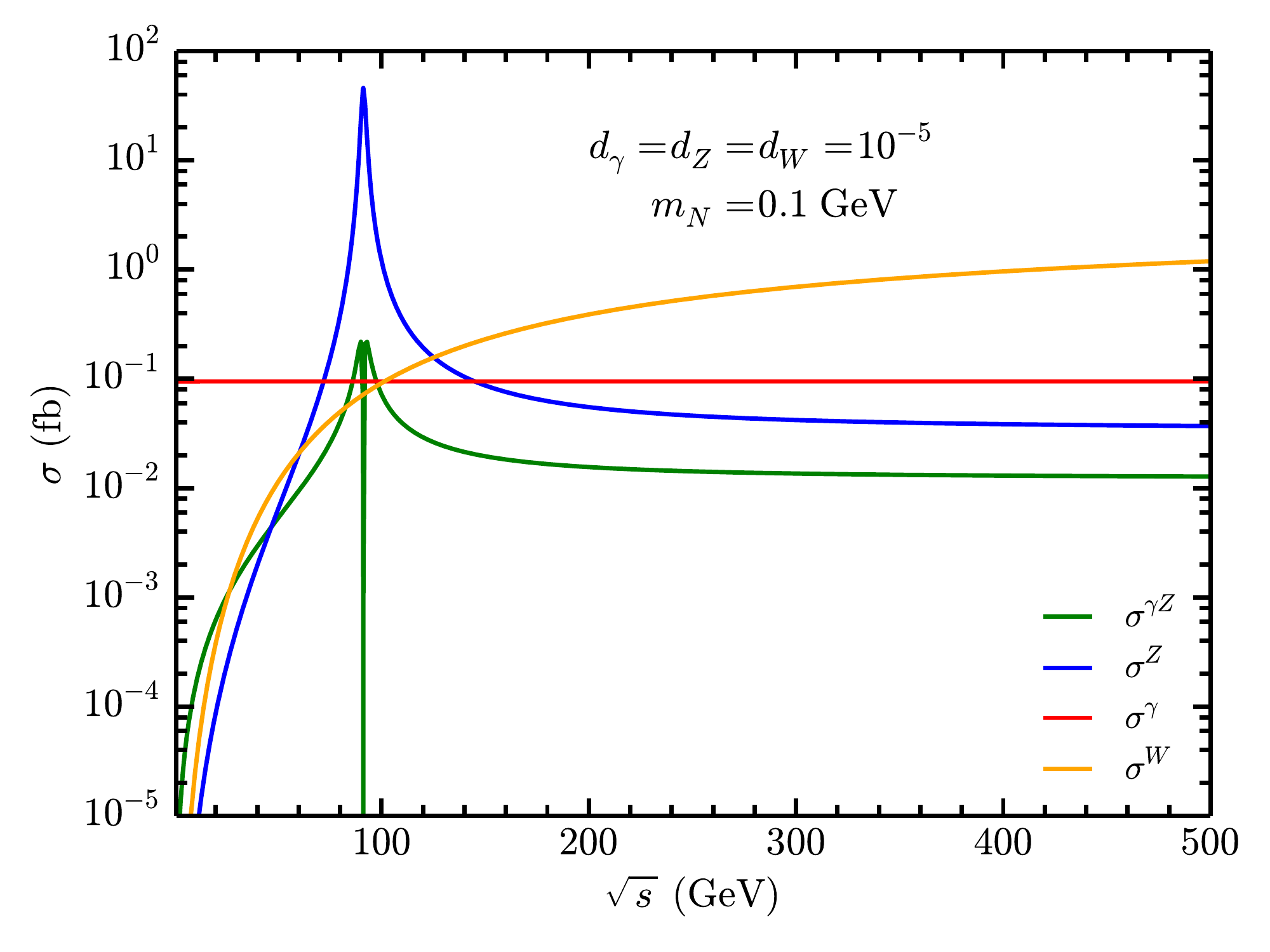}
 		\includegraphics[width=0.45\columnwidth]{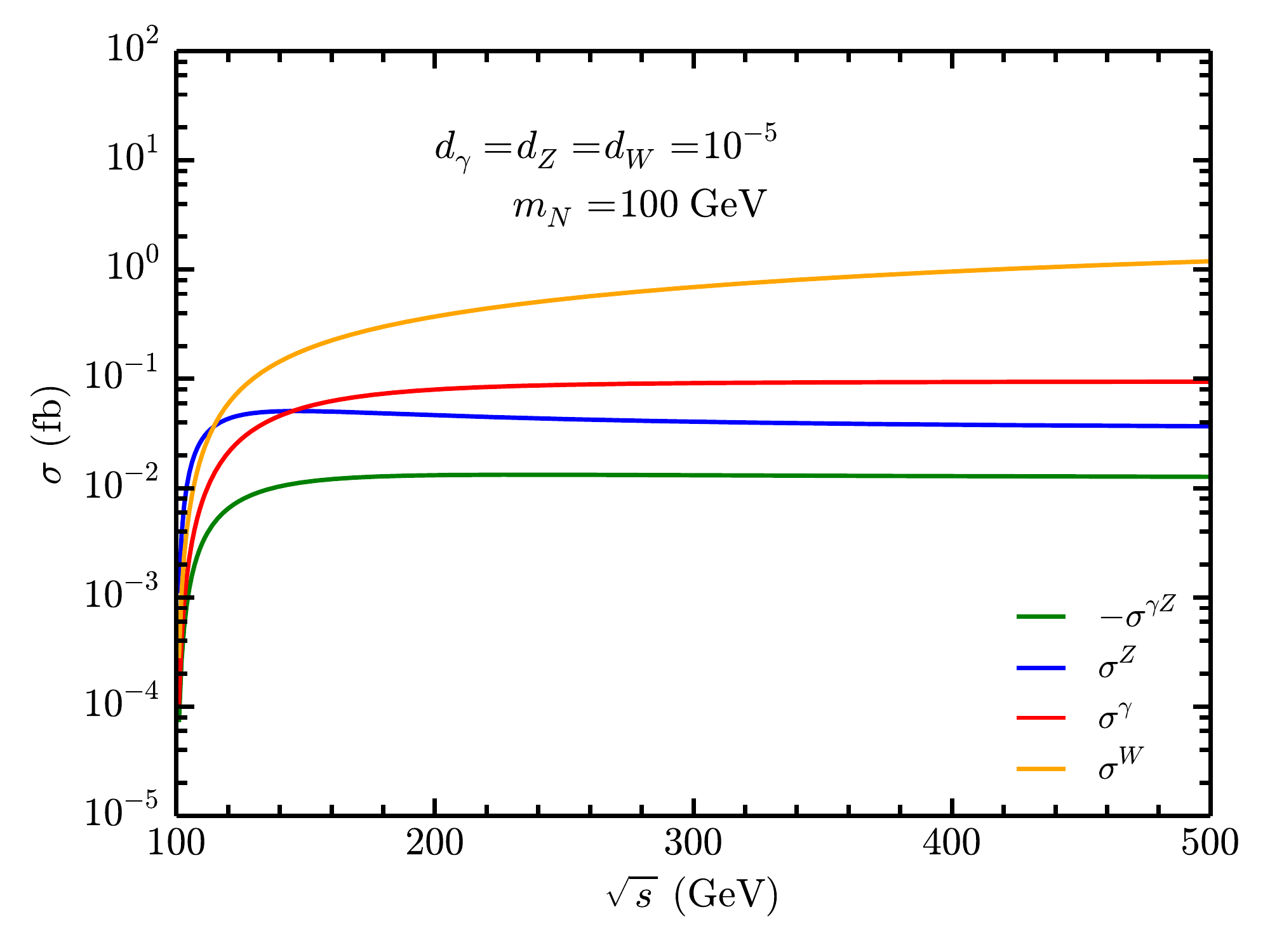}
 		\caption{The cross sections of the HNL associated with electron neutrino production as the function of CM energy  $\sqrt{s}$ for $m_N= 0.1$ GeV (left) and $m_N= 100$  GeV (right)  from $\gamma$, $Z$ and $W$ mediators with $d_\gamma=d_Z=d_W=10^{-5}$. Since the cross section $\sigma^{\gamma Z}/(d_\gamma d_Z)<0$ when $\sqrt{s}>M_Z$, its absolute value is shown. }
 		\label{fig:sigma}
 	\end{centering}
 \end{figure}

 In Fig. \ref{fig:sigma}, we present the cross sections of the HNL associated with electron neutrino production as the function of CM energy  $\sqrt{s}$ for $m_N= 0.1$ GeV (left) and $m_N= 100$  GeV (right)  from $\gamma$, $Z$ and $W$ mediators with $d_\gamma=d_Z=d_W=10^{-5}$, separately. Noted that, when $\sqrt{s}>M_Z$, there will be $\sigma^{\gamma Z}/(d_\gamma d_Z)<0$, thus absolute values of $\sigma^{\gamma Z}$ are plotted in Fig. \ref{fig:sigma}. 
 We can see that the contribution   from $\gamma$ mediator $\sigma^{\gamma}$ has little to do with the CM energy when $m_N\ll \sqrt{s}$. The contribution $\sigma^{Z}$ from $Z$ mediator for $m_N= 0.1$ GeV reaches its maximum when $\sqrt{s}=M_Z$  due to the $Z$ resonance. The contribution $\sigma^{W}$ from $W$ mediator only appears in $N\bar\nu_e$ production, and can be ignored comparing with $\sigma^{Z}$ when $\sqrt{s}$  around $Z$-pole. While  $\sigma^{W}$ always increase with the increment of $\sqrt{s}$, and will be dominant when $\sqrt{s}\gg M_Z$. 
 Just thanks to the additional contribution $\sigma^{W}$ to electron neutrino production, the sensitivity on dipole coupling $d_\gamma^e$ will be different from $d_\gamma^\mu$ and $d_\gamma^\tau$ at electron colliders when $\sqrt{s} >M_Z$.
 	
 %%%%%%%%%%%%%%%%%%%%%%%%%%%%%%%%%%%%%%%%%%%%%%%%%%%%%%%%%%%%%%%
 %%%%%%%%%%%%%%%%%%%%%%%%%%%%%%%%%%%%%%%%%%%%%%%%%%%%%%%%%%%%%%%

	To make sure that there exists visible photon in the final state, the subsequent decay of $N$ must occur inside the fiducial volume of the detector. 
	The probability of the heavy neutrino to
	decay radiatively in the fiducial volume after traveling a distance $l$ from
	the primary vertex is given by
	\begin{equation}
		P_{dec}(l)=(1-e^{-l/l_{dec}}){\rm Br}(N\to\nu\gamma).
	\end{equation}
	The decay length of $N$, $l_{dec}$, scales as 
	\begin{equation}
	l_{dec}=c\tau\beta\gamma=\frac{4\pi}{d_\gamma^2m_N^4}\sqrt{E_N^2-m_N^2}
	\end{equation}	
	in the case of ${\rm Br}(N\to\nu\gamma)\simeq 1$,
	where $E_N$ is the energy of $N$, with $E_N=\frac{s+m_N^2}{2\sqrt{s}}$ in the
	process $e^+e^-\to  N\bar\nu$.
	
	Then, the production rates from new physics in signal can be given as 
	\begin{equation}
			\sigma^{\rm NP}(e^+e^-\to\gamma+{\rm INV})=\sigma(e^+e^-\to N\nu){\rm Br}(N\to\nu\gamma)
		\epsilon_{cuts}\epsilon_{det}P_{dec}(l_D),
	\end{equation}	
	where $l_D$ is the detector length, $\epsilon_{cuts}$
	and $\epsilon_{det}$ are the efficiencies of the kinematic cuts and detection for the final photon, respectively.
	Since $N$ is usually produced on-shell and travels some distance before decaying,
	we employ the narrow width approximation to derive the kinematic imformation
	of the final state photon.
	The $1-\cos\theta$ distribution is used for the photon from 
	$N$ decay, where $\theta$ is the photon angle in the rest frame of $N$ \cite{Li:1981um,Masip:2012ke}.

	\section{Constraints from LEP}\label{sec:LEP}
	
	There are luxuriant  analyses to search monophoton signature at LEP, which can be used to set the constraints on dipole portal coupling to HNLs.
	In this section, we consider the single photon events with missing energy at the  CM energies around  $Z$  pole at LEP1 and 
	at larger CM energies above  $Z$  pole at LEP2. If the coupling between HNL and $Z$ boson exists, there will be addition constraints from $Z$ decay, which has been measured accurately by LEP.

	\subsection{LEP1}	
For the CM energies around  $Z$  pole at LEP1, 
the 95\% confidence level (C.L.) upper limits on the integrated cross section for  production of a single photon with $E_\gamma>E_{\rm min}$   and $|\cos\theta_\gamma| < 0.7$ are presented as the function of a specified minimum energy $E_{\rm min}$  by the OPAL Collaboration \cite{OPAL:1994kgw}.
We adopt the 95\% C.L. limit of 0.15 pb on the cross section for production of a single photon with energy exceeding $E_{\rm min}= 23$ GeV in the $|\cos\theta_\gamma| < 0.7$ angular region to give the corresponding  95\% C.L. limit on  the dipole portal to HNLs. The corresponding upper bounds on the dipole coupling $d_\gamma$ are shown in Fig. \ref{fig:LEP} with green lines for the four scenarios listed in Table \ref{tab:scen}, respectively. The overall detection efficiency of photon is estimated to be $65.7$\% \cite{OPAL:1994kgw}.

	\subsection{LEP2}
For larger CM energies above  $Z$  pole at LEP2, we use the DELPHI data of single photon at $\sqrt{s}=200\sim209$ GeV (average $\sqrt{s}=205.4$ GeV) in the angular region of  $45^{\circ}<\theta_\gamma<135^{\circ}$ and $0.06<x_\gamma<1.1$ with $x_\gamma=2 E_\gamma/\sqrt{s}$ \cite{DELPHI:2003dlq}.
The constraints on the dipole coupling can be obtained by performing a simple $\chi^2$ analysis with
the function of 
\begin{equation}
	\chi^2=\left(\frac{\sigma^{\rm SM}+\sigma^{N\nu}-\sigma^{\rm exp}}{\delta\sigma^{\rm exp}}\right)^2,
\end{equation}
with $\sigma^{\rm SM}=1.61$ pb,  $\sigma^{\rm exp}=1.50$ pb, and $\delta\sigma^{\rm exp}=0.11$ pb from  Ref. \cite{DELPHI:2003dlq}.
%Then the 95\% C.L. upper limit on the cross section of new physics is found to be about 0.11 pb.
Estimated from the Monte Carlo cross sections and the expected numbers of events, the  overall detection efficiency of photon is set to be 65\%.
The 95\% C.L. upper limits on the dipole coupling $d_\gamma$ are shown in Fig.\ref{fig:LEP} with blue lines  for the four scenarios.

	\subsection{$Z$ decay}
\begin{figure}[htbp]
	\begin{centering}
		\includegraphics[width=0.45\columnwidth]{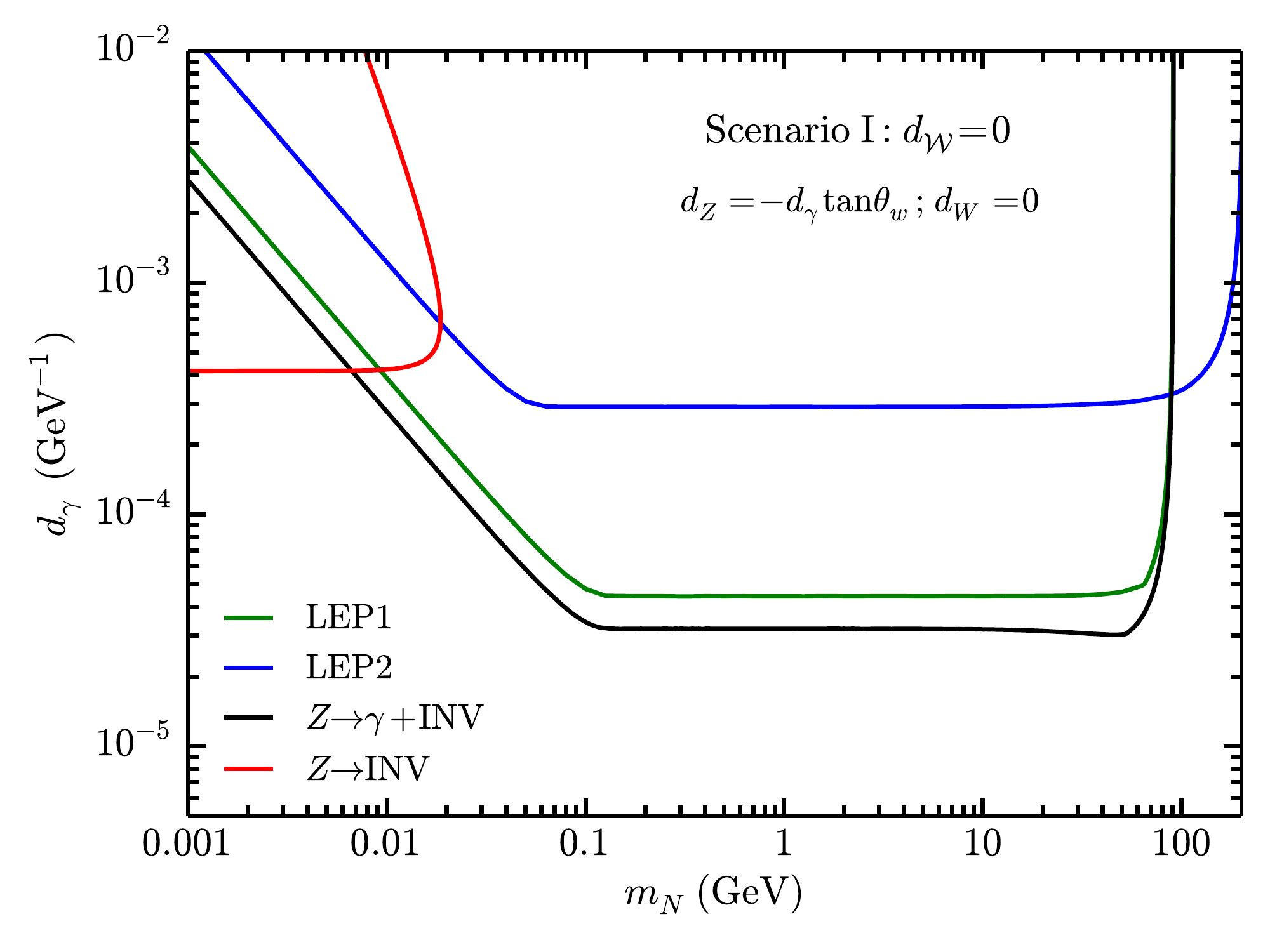}
		\includegraphics[width=0.45\columnwidth]{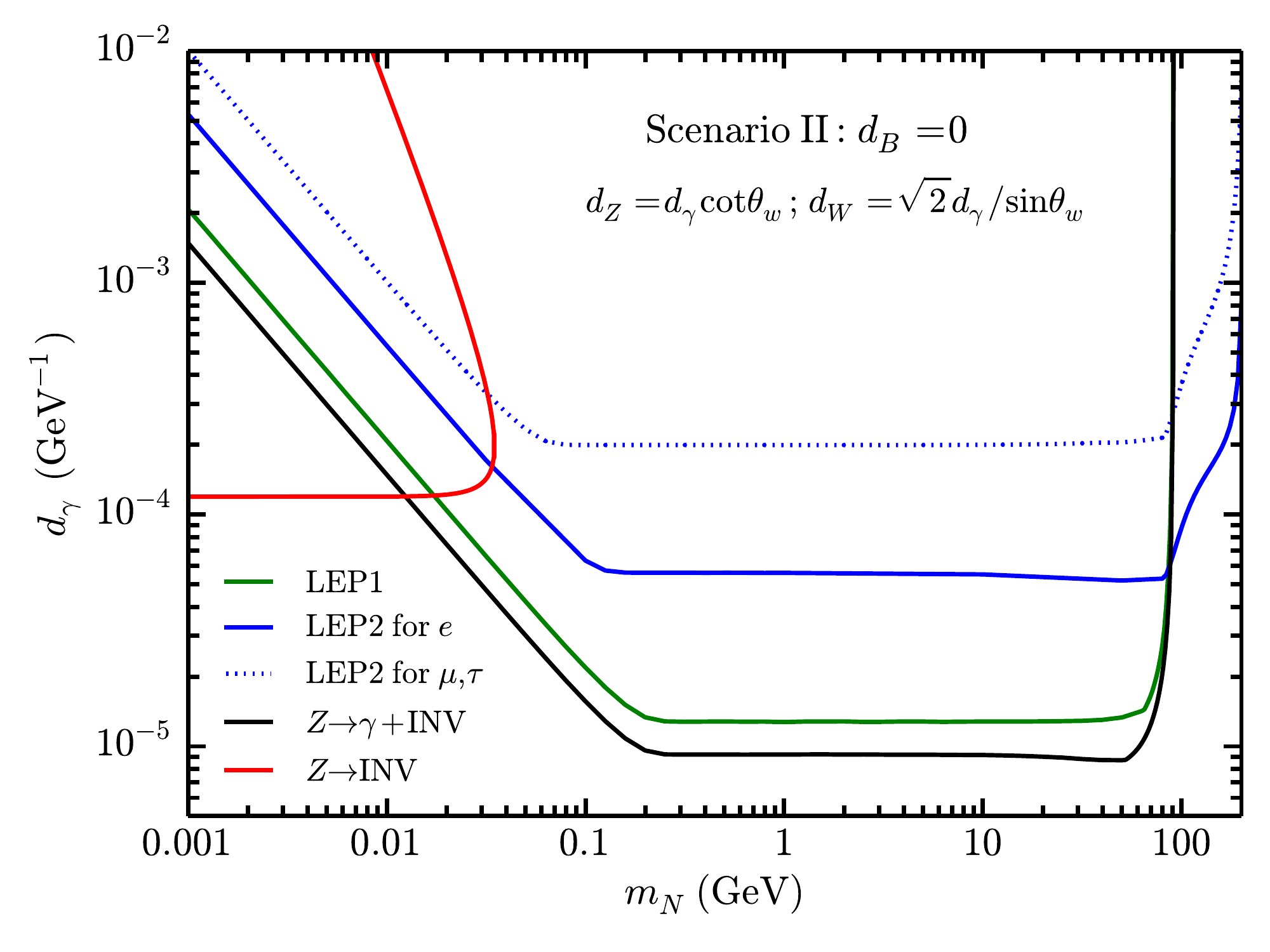}
		\includegraphics[width=0.45\columnwidth]{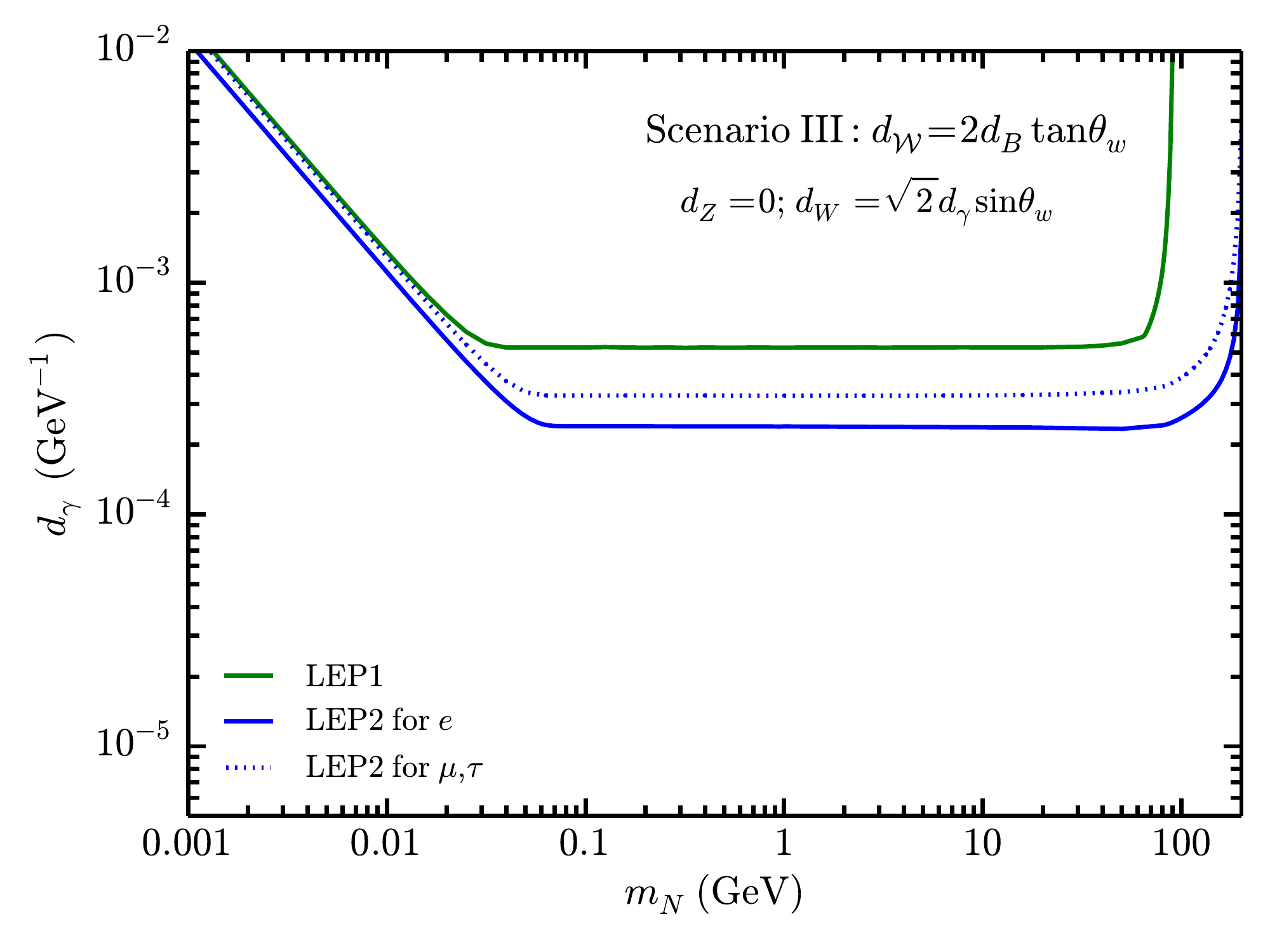}
		\includegraphics[width=0.45\columnwidth]{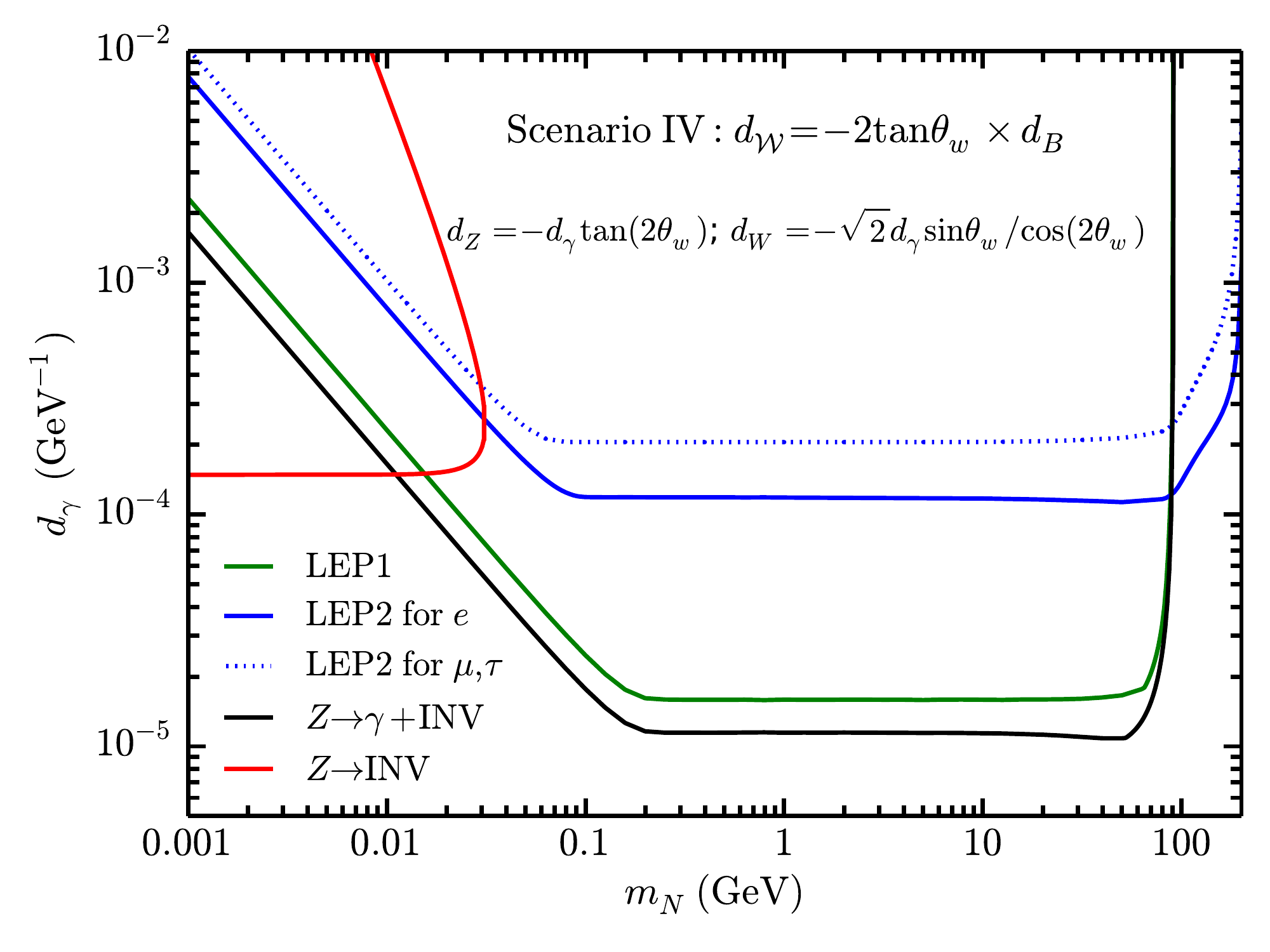}	
		\caption{The 95\% CL upper bounds on the dipole portal coupling $d_\gamma$ under  four assumptions listed in Table \ref{tab:scen}  from the monophoton searches at LEP1 \cite{OPAL:1994kgw} (green lines) and LEP2 \cite{DELPHI:2003dlq} (blue solid lines for $d_\gamma^e$ and blue dotted  lines for $d_\gamma^{\mu,\tau}$), the decay $Z\to\gamma+\rm{invisible}$ \cite{L3:1997exg} (black lines) and $Z$ invisible decay  \cite{ALEPH:2005ab} (red lines), respectively. }
		\label{fig:LEP}
	\end{centering}
\end{figure}

Negative evidence for the single photon with missing energy signal at L3 detector of LEP1 \cite{L3:1997exg}, set an upper limit at the 95\% C.L. lying in the range of  about $(3.2\sim1.1) \times 10^{-6}$ on the branching ratio for $Z$ decaying to invisible particles and a photon with  energy greater that $E_{\rm min}$ in the range of $(15\sim40)$ GeV. 
The measurable decay width $\Gamma_{Z\to\gamma+{\rm invisible}}$ at LEP from dipole portal to HNLs can be expressed as 
\begin{equation}
	\Gamma_{Z\to\gamma+{\rm invisible}}^{\rm NP}=(\Gamma_{Z\to N\bar\nu}+\Gamma_{Z\to \bar N \nu}){\rm Br}(N\to\nu\gamma)\epsilon_{cuts}(1-P_{dec}(l_D)).
\end{equation}
Here the decay width of $Z\to N\bar\nu$ or $Z\to \bar N \nu$ can be given as
\begin{equation}
\Gamma_{Z\to N\bar\nu}=\Gamma_{Z\to \bar N \nu}=\frac{d_Z^2(M_Z^2-m_N^2)^2(2m_N^2+M_Z^2)}{12\pi m_Z^3} \Theta(m_Z>M_N),
\end{equation}
and $P_{dec}(l_D)$ denotes the probability of  heavy neutrino $N$ to
decay radiatively out the detector at LEP with the detector length $l_D=1$ m,
$\epsilon_{cuts}$ is the efficiency of the kinematic cuts with $E_\gamma > E_{\rm min}$ for the final photon.
We use the 95\% C.L. upper limit of $3.2 \times 10^{-6}$ on the branching ratio ${\rm Br}(Z\to\gamma+{\rm invisible})$ with $E_\gamma > 15$ GeV to provide the corresponding 95\% C.L. constraint on dipole portal coupling $d_\gamma$, which is presented in Fig. \ref{fig:LEP} with black lines. 

Besides,  $N$ decaying out of the detector will contribute to the $Z$-boson invisible decay as
\begin{equation}
	\Gamma^{\rm NP}_{Z\to{\rm invisible}}=(\Gamma_{Z\to N\bar\nu}+\Gamma_{Z\to \bar N \nu})P_{dec}(l_D).
\end{equation}
The total width of the $Z$ boson has been measured accurately by the LEP experiments which place a strong bound on new physics contributions $\Gamma^{\rm NP}_{Z\to{\rm invisible}}< 2.0$ MeV at 95\% C.L. \cite{ALEPH:2005ab}.
The 95\% C.L. upper limits from $Z$ invisible decay on the dipole coupling are given in Fig. \ref{fig:LEP} with red lines.

%%%%%%%%%%%%%%%%%%%%%%%%%%%%%%%%%%%%%%%%%%%%%%%%%%%%%%%%%%%%%%%
%%%%%%%%%%%%%%%%%%%%%%%%%%%%%%%%%%%%%%%%%%%%%%%%%%%%%%%%%%%%%%%

\subsection{Results}

One sees that the constraints from monophoton searches at LEP1 and LEP2, and from $Z$ decaying into invisible particles associated with a photon have a characteristic ``U" shape. The right boundary of the
 ``U" shape region for lager $m_N$ is controlled by the kinematic reach,
and in the case of the LEP2 extends beyond 100 GeV.
The left boundary of the excluded  ``U" shape region for small $m_N$, is   controlled by the lifetime of $N$.
Since smaller $m_N$ leads to the longer lifetime of $N$, $N$ will more likely decay out of the detector with the loss of the $\gamma$ signal. The $Z$ invisible decay  can provide complementary constraints for the HNLs with small mass.

Noted that because of the additional contribution from $W$ mediator diagram in $t$-channel for $N\nu_e$ production , the constraints on $d_\gamma^e$ (blue solid lines) will be stricter than $d_\gamma^{\mu,\tau}$ (blue dotted lines) from monophoton searches at LEP2 with $\sqrt{s} > M_Z$ when $d_W\neq0$, which can be seen in scenarios II, III, IV. In scenarios I, II and IV with $d_Z\neq 0$, there will be additional constraints from $Z$ decay. The measurements of $Z$ decay will derive same sensitivity for all the three  lepton flavors, so almost do the monophoton searches at LEP1, since $\sigma^W$ can be ignored comparing with  $\sigma^Z$  around $Z$-pole.

Since LEP1 with $\sqrt{s}\simeq M_Z$ can provide very competitive production rates of HNL  due to the $Z$-resonance from $Z$-mediator in $s$-channel when $d_Z\neq0$,  the sensitivities on the dipole portal coupling $d_\gamma$ are much better at LEP1,
which can be about one order of magnitude,  than the ones at LEP2 with $m_N\lesssim90$ GeV in scenarios I, II and IV.
While in scenario III with $d_Z=0$, LEP2 always give leading constraints in all the plotted mass region.
Though there are no $Z$-resonance enhancement in scenario III with $d_Z=0$ at LEP1, the limits for different lepton flavors are still almost the same since $\sigma^\gamma$ is more dominant than $\sigma^W$ around $Z$-pole.
  The constraints from the measurement of  the branching ratio for $Z\to\gamma+\rm{invisible}$ are always found to be more stringent than  the ones from monophoton searches at LEP1 in the scenarios with $d_Z\neq0$.

\begin{figure}[htbp]
	\begin{centering}
		\includegraphics[width=0.45\columnwidth]{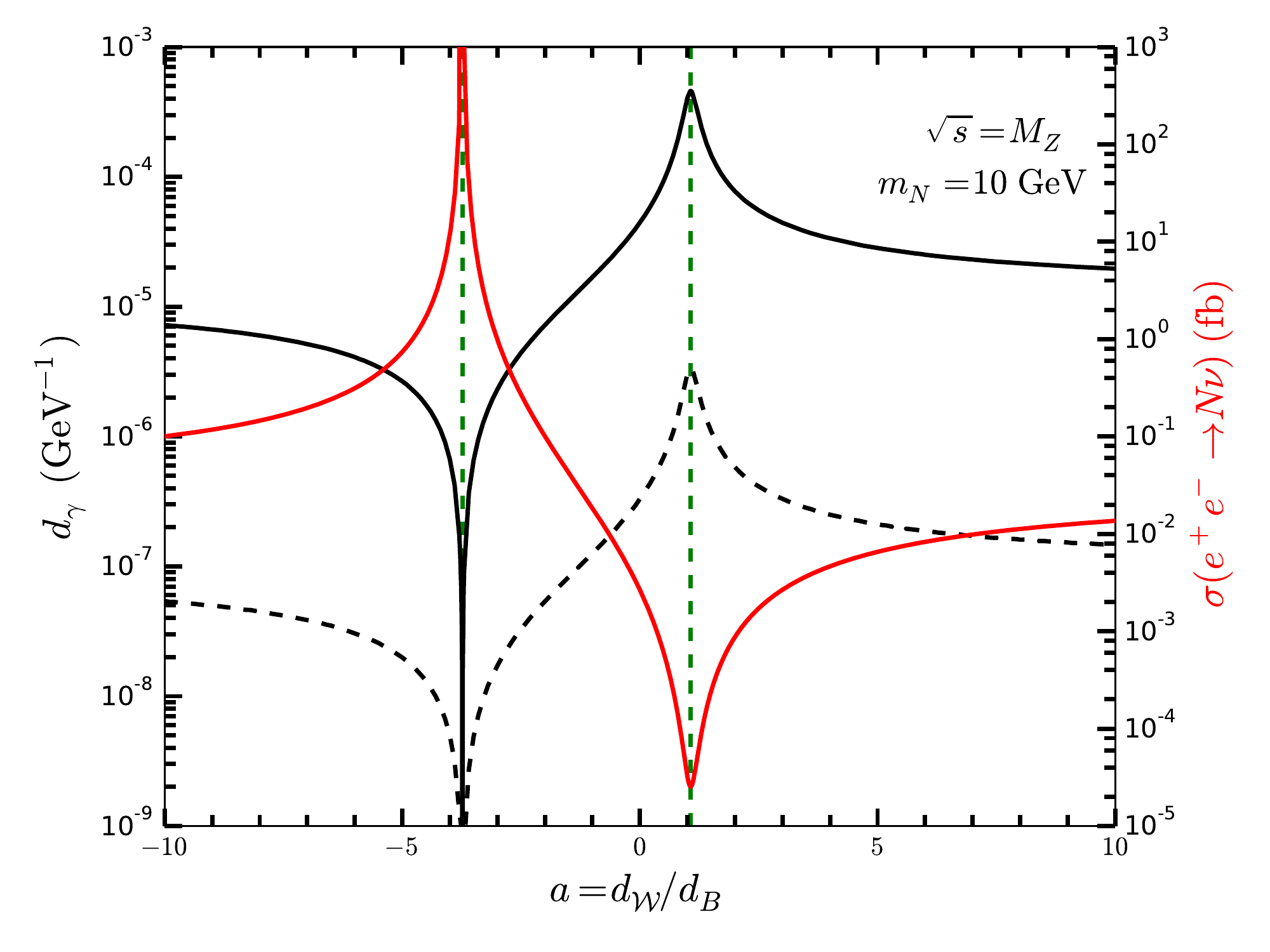}
		\includegraphics[width=0.45\columnwidth]{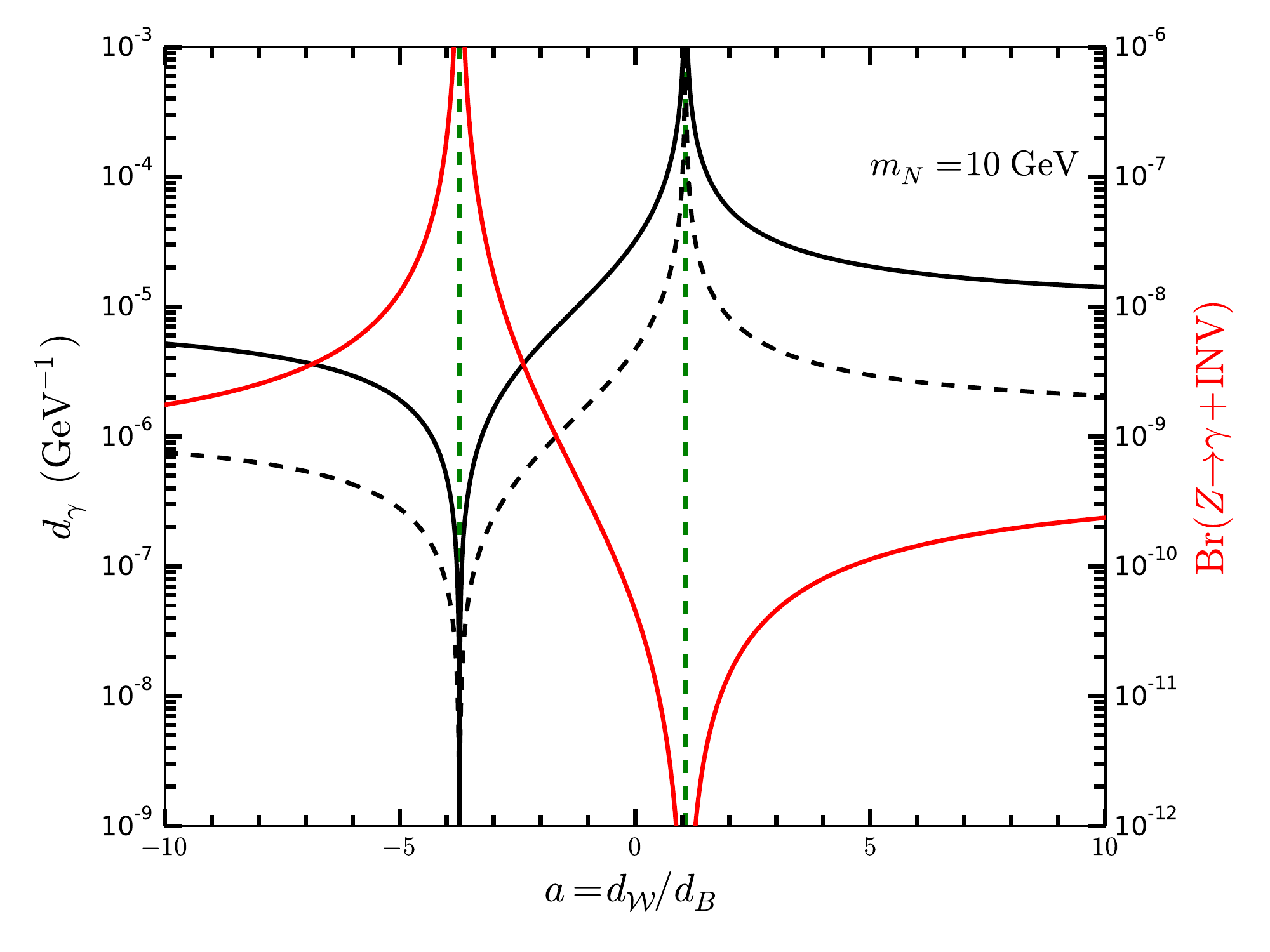}
		\caption{Left: The production rates for the process $e^+e^-\to N \bar\nu$ (red line) with $m_N=10 $ GeV, $d_\gamma=10^{-7}$ and $\sqrt{s}=M_Z$ , and the 95\% C.L. upper limits on the neutrino dipole portal coupling to HNLs  $d_\gamma$ as the function of the ratio $a=d_\calW/d_B$ with the CM energy on $Z$-pole using the monophoton searches by the OPAL Collaboration at LEP1 \cite{OPAL:1994kgw}  (black solid line) and future CEPC (black dashed line) with the luminosity of 100 ab$^{-1}$. Right: The branching ratio for $Z$ decaying into invisible particles and a photon ${\rm Br}(Z\to\gamma+{\rm invisible})$ (red line) with $m_N=10 $  GeV and $d_\gamma=10^{-7}$, and the constraints on $d_\gamma$ as the function of the ratio $a=d_\calW/d_B$ with the assuming  ${\rm Br}(Z\to\gamma+{\rm invisible})=10^{-7}$ in the future (black dashed line) and 95\% C.L. upper upper limit of ${\rm Br}(Z\to\gamma+{\rm invisible})=3.2 \times 10^{-6}$ with $E_\gamma > 15$ GeV  from LEP (black solid line).  }
		\label{fig:dwitha}
	\end{centering}
\end{figure}

In Fig. \ref{fig:dwitha}, we present the production rate for the sterile neutrino associated with active neutrino at electron colliders with $\sqrt{s}=M_Z$ (left), and the branching ratio of $Z\to \gamma+{\rm invisible}$ (right), as the function of $a=d_\calW/d_B$, which are all labeled with red line, respectively.
Here we set $m_N=10$ GeV and $d_\gamma=10^{-7}$. 
Since there is a singularity in $d_Z$ and $d_W$ at $a=-2\cot\theta_{w}\sim -3.7$, the production rate of $N\nu$ and the branching ratio of $Z\to \gamma+{\rm invisible}$ will increase when $a < -2\cot\theta_{w}$ with the increment of $a$, then decrease until $a=2\tan\theta_{w}\sim 1.1$.
With $a=2\tan\theta_{w}$, the dipole coupling with $Z$ boson $d_Z$ becomes zero, the production rate reaches its minimum and the branching ratio goes to zero.
The corresponding 95\% C.L. upper limits on dipole coupling $d_\gamma$ as the function of $a$ from the monophoton searches (left) and the measurement of   $Z\to \gamma+{\rm invisible}$ (right) at LEP1 are also shown in Fig. \ref{fig:dwitha} with black solid lines, respectively.
In the case of $a>0$, the upper bounds on $d_\gamma$ lie in the range of $(2.0\times 10^{-5}, 4.5\times10^{-4})$ from monophoton searches at LEP1.
Besides the region very near $a=2\tan\theta_{w}$ where $d_Z$ is close to zero, the measurement of $Z$ decaying into photon plus invisible particles can provide stricter constraint than monophoton searches at LEP1.

%%%%%%%%%%%%%%%%%%%%%%%%%%%%%%%%%%%%%%%%%%%%%%%%%%%%%%%%%%%%%%%
%%%%%%%%%%%%%%%%%%%%%%%%%%%%%%%%%%%%%%%%%%%%%%%%%%%%%%%%%%%%%%%

\begin{figure}[htbp]
	\begin{centering}		
		\includegraphics[width=0.45\columnwidth]{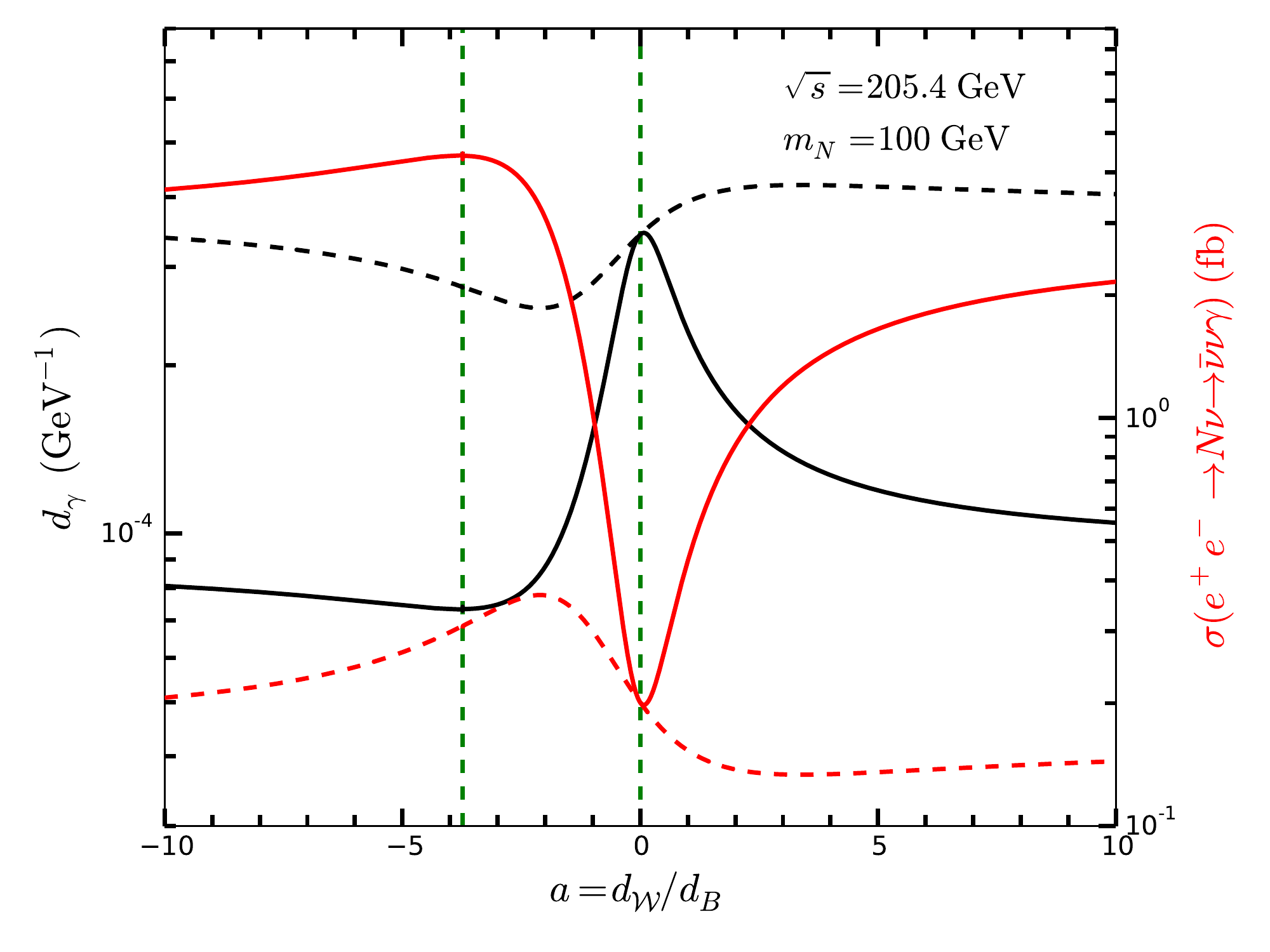}
		\includegraphics[width=0.45\columnwidth]{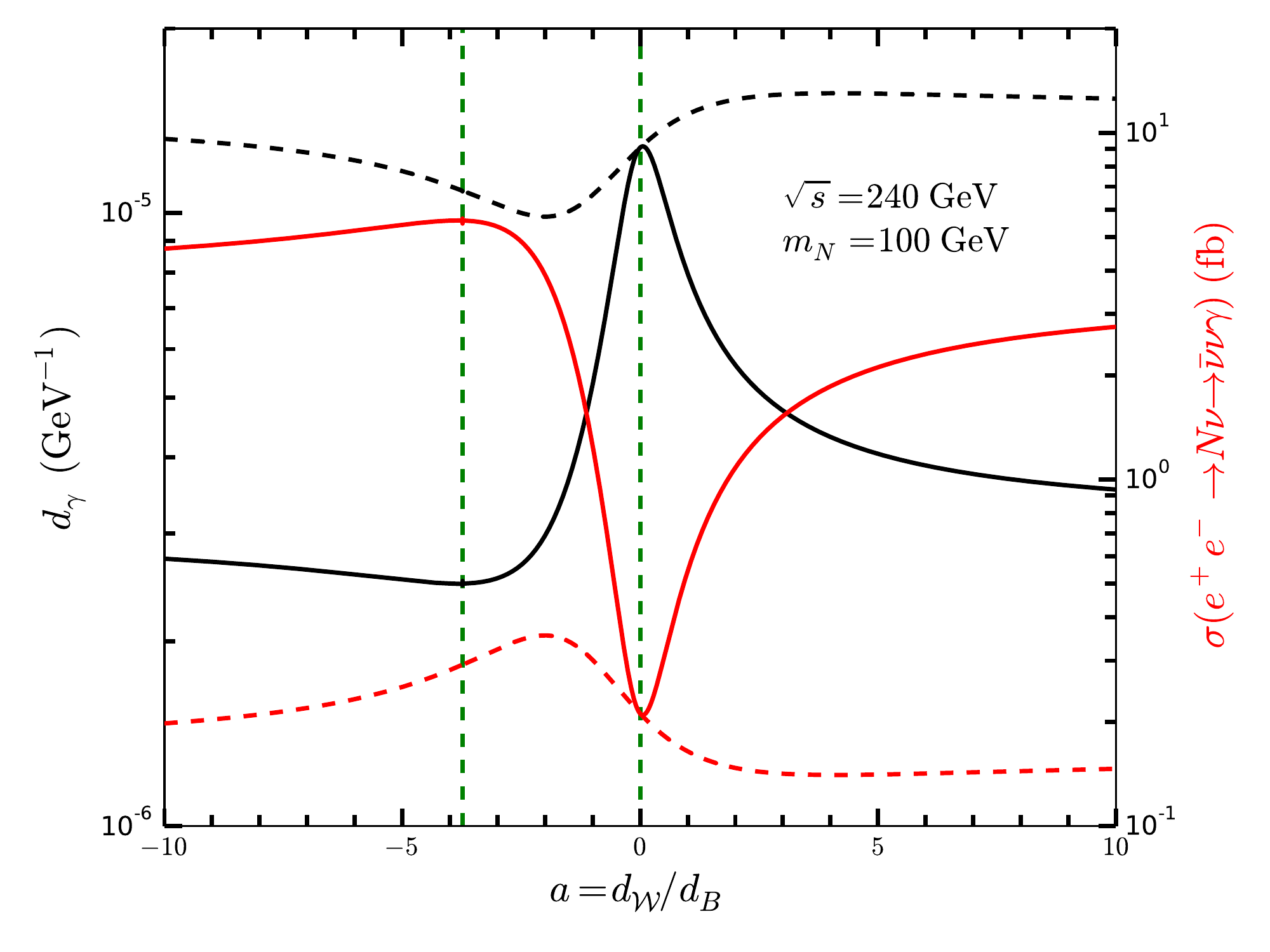}
		\caption{The production rates for the processes $e^+e^-\to N\nu_e\to  \nu_e\bar\nu_e\gamma$ (red solid line) and $e^+e^-\to N\nu_{\mu,\tau}\to  \nu_{\mu,\tau}\bar\nu_{\mu,\tau}\gamma$ (red dashed line) with $m_N=100$ GeV, $d_\gamma=10^{-5}$, and $\sqrt{s}=205.4$ GeV, and the 95\% C.L. upper limits on the neutrino dipole portal couplings $d_\gamma^e$ (black solid line) and $d_\gamma^{\mu,\tau}$  (black dashed line) as the function of the ratio $a=d_\calW/d_B$ using the monophoton searches  by the OPAL Collaboration at LEP2  \cite{OPAL:1994kgw} (left) and future CEPC in $H$-mode with the luminosity of 20 ab$^{-1}$ (right).  }
		\label{fig:dwitha100}
	\end{centering}
\end{figure}

The graph on the left of Fig. \ref{fig:dwitha100} shows the production rates of the  processes $e^+e^-\to N\nu_e\to  \nu_e\bar\nu_e\gamma$ (red solid line) and $e^+e^-\to N\nu_{\mu,\tau}\to  \nu_{\mu,\tau}\bar\nu_{\mu,\tau}\gamma$ (red dashed line) as the function of $a$ with $m_N=100\ \GeV$ and $d_\gamma=10^{-5}$ at LEP2 with $\sqrt{s}=205.4$ GeV. The  kinematic cuts and the detection efficiency  for the final photon in DELPHI data \cite{DELPHI:2003dlq}  are also considered.
Interestingly, the cross sections of monophoton due to $N$ production for $m_N=100$ GeV at electron collider  are observed to change not too much around $a=-2\cot\theta_{w}$, not as drastically as in Fig. \ref{fig:dwitha} for  $m_N=10$ GeV.
This is because that when $m_N>M_Z$, the openning of decay channel $N\to \ell W$ and $N\to \nu Z$ will reduce the branching ratio of $N\to \nu\gamma$, which is inversely proportional to $(d_Z^2+d_W^2)$ and   thereby offsets the dependence of the production rate for monophoton on the couplings $d_Z$ or $d_W$.
From the different between the production rates of $ \nu_e\bar\nu_e\gamma$  and $\nu_{\mu,\tau}\bar\nu_{\mu,\tau}\gamma$ in Fig. \ref{fig:dwitha100}, one can find the additional contribution from $W$-mediator diagram in $t$-channel for the production of $N$.
The corresponding  95\% C.L. upper limits on the dipole portal couplings $d_\gamma^e$ (black solid line) and $d_\gamma^{\mu,\tau}$  (black dashed line) for $m_N=100$ GeV using DELPHI data of the monophoton search  \cite{DELPHI:2003dlq} at LEP2 are also shown in the graph on the left of Fig. \ref{fig:dwitha100}.
For $|a|\le 10$, the upper limits on the dipole portal couplings to HNL with mass of 100 GeV, $d_\gamma^e$ and $d_\gamma^{\mu,\tau}$, lie in the range of $(7.1\times10^{-5},3.5\times10^{-4})$ and $(2.5\times10^{-4},4.1\times10^{-4})$  from monophoton searches at LEP2, respectively. When $a=0$ ($d_W=0$), the upper limit on  $d_\gamma^e$ reaches its maximum.

%%%%%%%%%%%%%%%%%%%%%%%%%%%%%%%%%%%%%%%%%%%%%%%%%%%%%%%%%%%%%%%
%%%%%%%%%%%%%%%%%%%%%%%%%%%%%%%%%%%%%%%%%%%%%%%%%%%%%%%%%%%%%%%

	\section{Constraints from CEPC}

	In the following, we will investigate the sensitivity on the dipole portal coupling to HNLs at  the  future electron collider CEPC~\cite{CEPCStudyGroup:2018rmc, CEPCStudyGroup:2018ghi}.
	The CEPC, proposed by the Chinese high
	energy physics community in 2012, is designed to run primarily at a CM energy
	of 240 GeV as a Higgs factory ($H$-mode) with a total  luminosity of $ 20\ \mathrm{ab}^{-1}$ for ten years running \cite{CEPCPhysicsStudyGroup:2022uwl}. 
	In addition, on the $Z$-pole as a $Z$
	factory ($Z$-mode), it will also be operated with a total luminosity of $100\ \mathrm{ab}^{-1}$ for two years, perform a precise $WW$ threshold scan ($WW$-mode) with a total luminosity of $\sim 6\ \mathrm{ab}^{-1}$ for one year running at 
	$\sqrt{s} \sim$ $160\ \mathrm{GeV}$, and will be upgraded to a CM energy
	of 360 GeV, close to the $t\bar t$ threshold ($t\bar t$-mode) with a total luminosity of $\sim 1\ \mathrm{ab}^{-1}$ for five years \cite{CEPCPhysicsStudyGroup:2022uwl}.

		In the search for monophoton signature at CEPC, the backgrounds can be classified into two categories:  the irreducible background and the reducible background.  The irreducible background arises from the neutrino production associated with one photon in SM $e^+e^-\to\nu\bar\nu\gamma$.  The reducible background comes from
		any SM process with a single photon in the final state with all other visible particles  undetected due to limitations of the detector acceptance. 
		Such as the  radiative Bhabha scattering, $e^+e^-\to e^+e^-\gamma $ should be considered carefully, which has a huge cross section and can mimic the signal if both the final state electrons and positrons escape undetected, for example, through the beam pipes \cite{Bartels:2012ex, Habermehl:2020njb}. 
		
		For the monophoton signature at CEPC, we use the cuts for the final detected photon following the CEPC CDR~\cite{CEPCStudyGroup:2018ghi}: $|z_\gamma|<0.99$ and $E_\gamma > 0.1 $  GeV (hereafter the ``preselection cut"). 
	Due to the SM $Z$ boson, the irreducible background from the SM neutrino pair production $e^+e^-\to\nu\bar\nu\gamma$ exhibits a resonance in the monophoton energy spectrum which exhibits a peak at the photon energy $E_\gamma^Z={(s-M_Z^2)}/{2\sqrt{s}}$ with a full-width-at-half-maximum
	as $\Gamma_\gamma^Z=M_Z\Gamma_Z/\sqrt{s}$. To suppress the irreducible background contribution, we will veto the events within  $E_\gamma\in(E_\gamma^Z\pm 5\Gamma_\gamma^Z)$ in the monophoton energy spectrum \cite{Liu:2019ogn} (hereafter the ``Z resonance veto cut"). 	We apply the cut 
	\begin{equation}
	E_\gamma >E_\gamma^m(\theta_\gamma)= \frac{\sqrt{s}}{(1+{\sin\theta_\gamma}/{\sin\theta_b})},
	\label{eq:adv-cuts}
	\end{equation}
	on the final state photon to remove the main reducible background from the processes $e^+e^-\to e^+e^-\gamma$ and $e^+e^-\to\gamma\gamma\gamma$ following Ref. \cite{Liu:2019ogn}, where $\theta_b$ denotes
	the angle at the boundary of the sub-detectors with $\cos\theta_b=0.99$. 
	{We will collectively refer to ``preselection cut", ``Z resonance veto cut" and cut of (\ref{eq:adv-cuts}) in the list as the ``basic  cuts"	hereafter.	
	}

\begin{figure}[htbp]
	\begin{centering}
		\includegraphics[width=0.45\columnwidth]{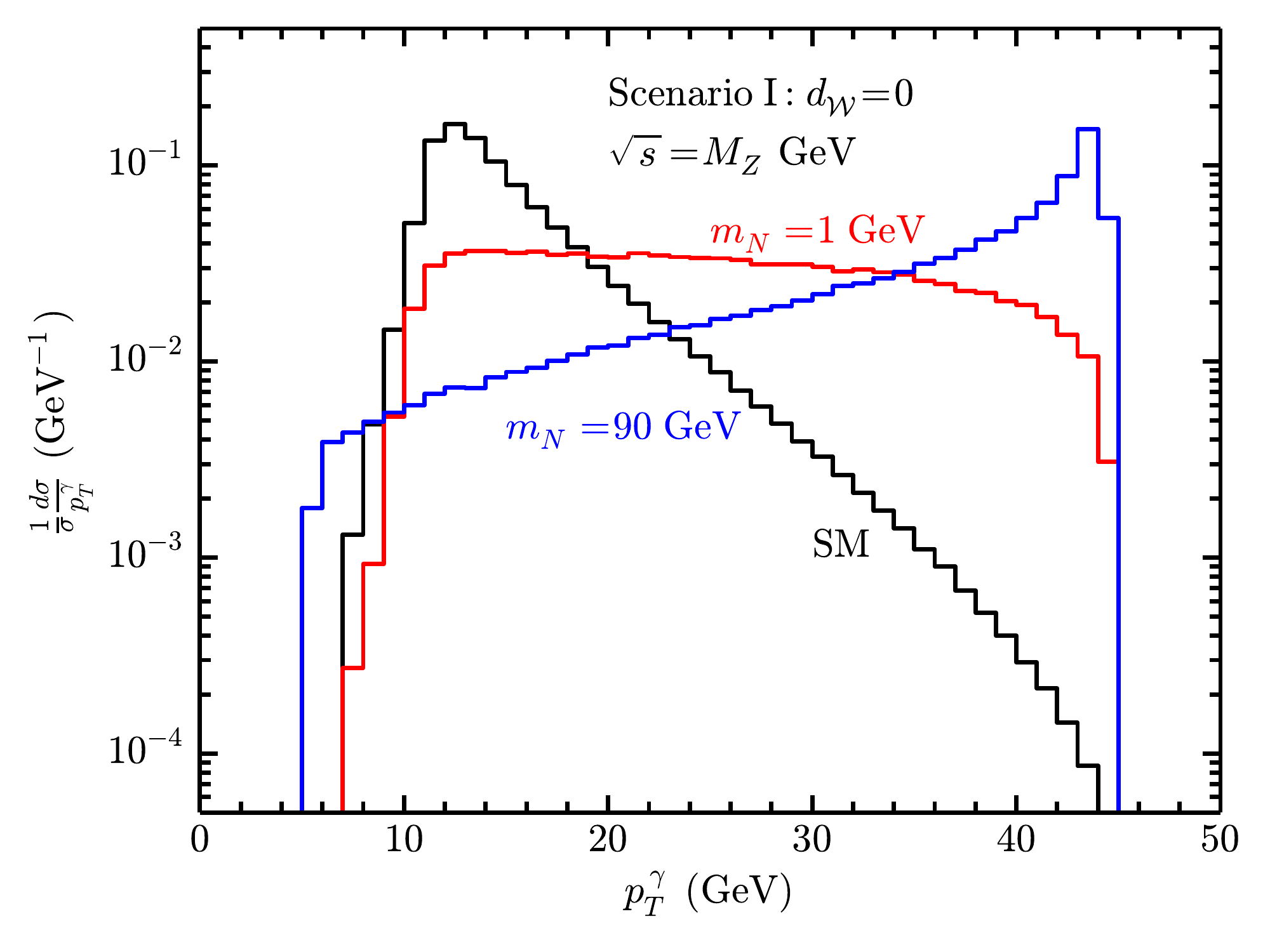}
		\includegraphics[width=0.45\columnwidth]{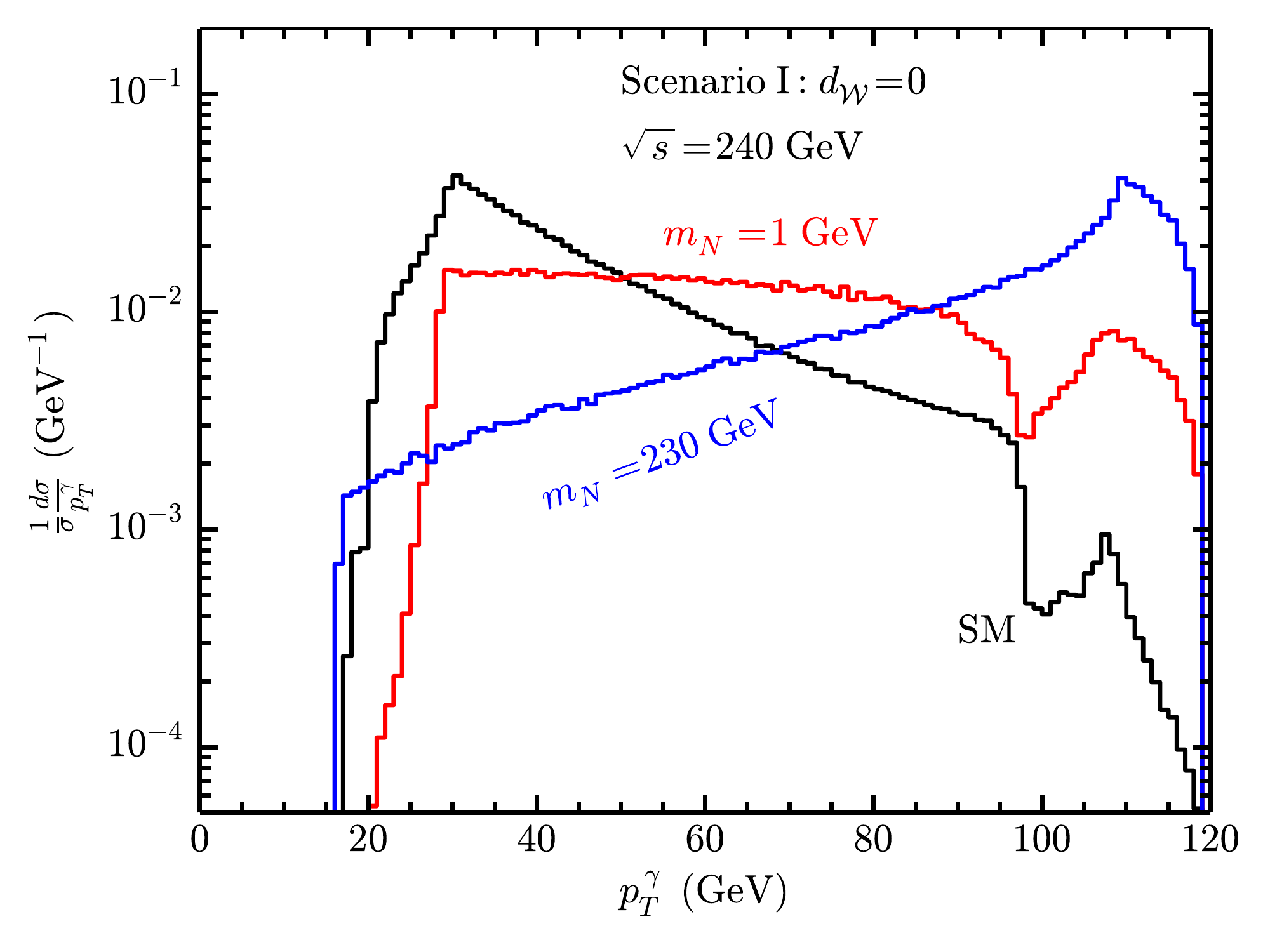}
		\caption{The normalized transverse momentum distribution of the final photon due to the background and the signal from  sterile neutrino after the ``advanced cut" at CEPC in $Z$-mode (left) and $H$-mode (right). For the signal from sterile neutrino, we consider $m_N=1$ GeV, and 90 GeV in the $Z$-mode, and $m_N=1$ GeV, and 230 GeV in the $H$-mode. }
		\label{fig:pt}
	\end{centering}
\end{figure}

	{In Fig. \ref{fig:pt}, we present the normalized  transverse momentum distribution of the final state photon due to the background and the signal from the sterile neutrino after the ``advanced cut" at CEPC in $Z$-mode and $H$-mode, respectively. It can be seen that, compared to the background, the typical feature of the signal  is that the final state photon are distributed in the large transverse momentum regions, especially for larger $m_N$. Thus, in order to improve the sensitivity, we impose the  transverse momentum cut in addition to the ``basic  cuts"	 for the final state photon
		with $p_T^\gamma > 35$ GeV in the $Z$-mode, $p_T^\gamma > 35$ GeV in the $Z$-mode, $p_T^\gamma > 65$ GeV in the $WW$-mode, $p_T^\gamma > 100$ GeV in the $H$-mode, and $p_T^\gamma > 160$ GeV in the $t\bar{t}$-mode, respectively. We collect the $p_T^\gamma$ cut and the ``basic detector cuts" as the the ``advanced  cuts". 
%		In Table \ref{tab:eff}, we show how much one can improve the sensitivity by using the  $p_T^\gamma$ cut.
	}

%%%%%%%%%%%%%%%%%%%%%%%%%%%%%%%%%%%%%%%%%%%%%%%%%%%%%%%%%%%%%%%
%%%%%%%%%%%%%%%%%%%%%%%%%%%%%%%%%%%%%%%%%%%%%%%%%%%%%%%%%%%%%%%

\begin{figure}[htbp]
		\begin{centering}
			\includegraphics[width=0.45\columnwidth]{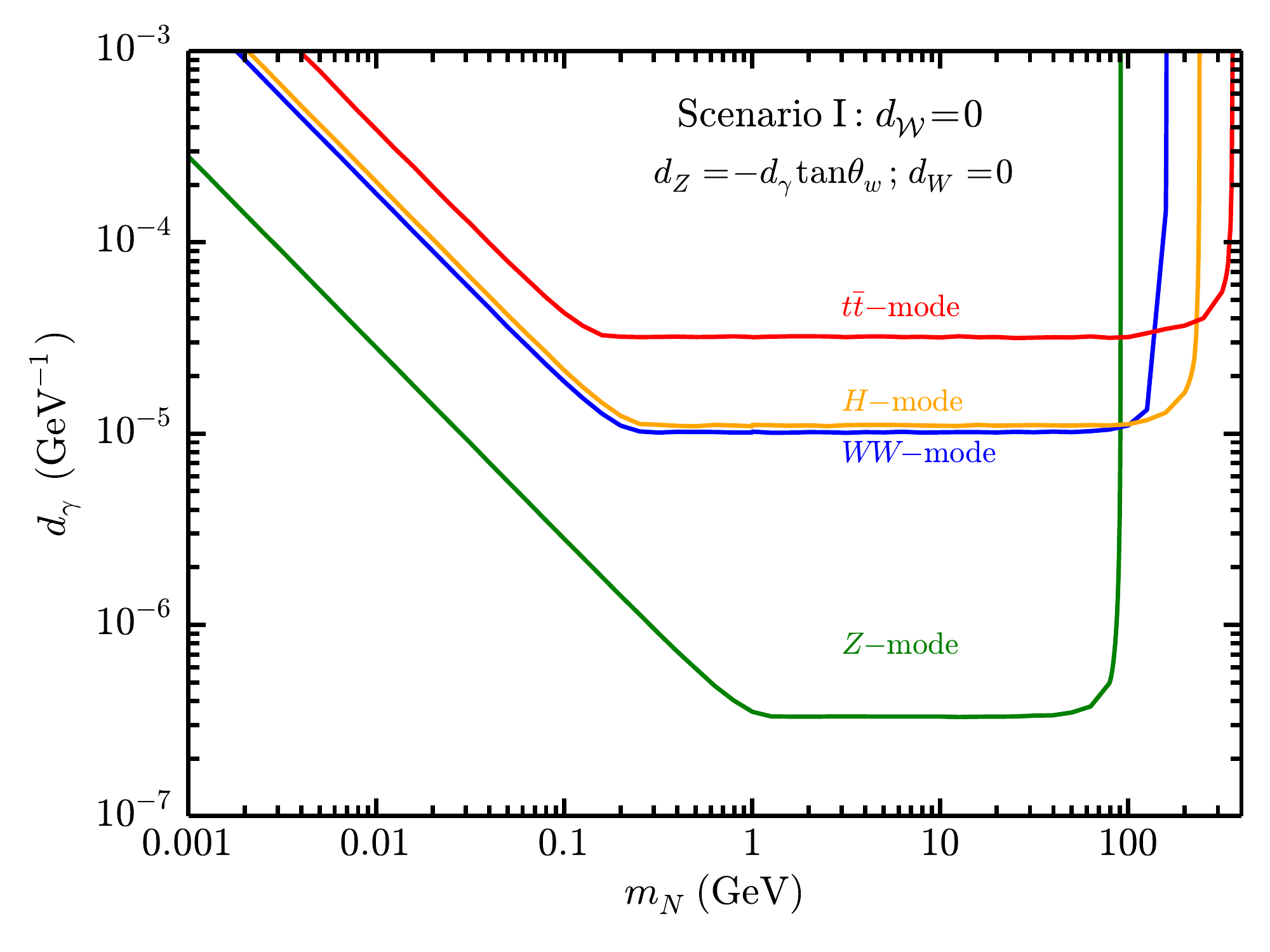}
			\includegraphics[width=0.45\columnwidth]{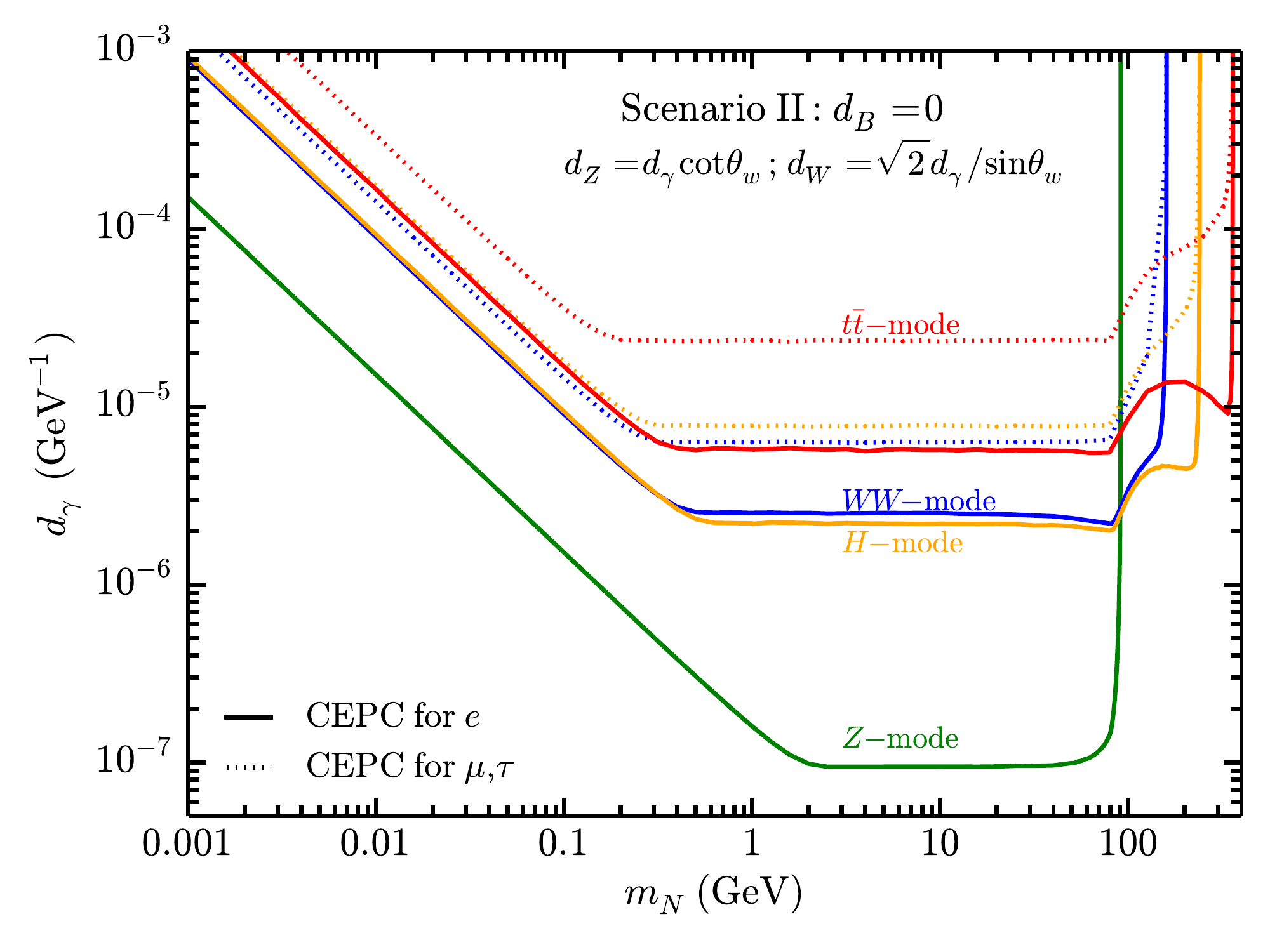}
			\includegraphics[width=0.45\columnwidth]{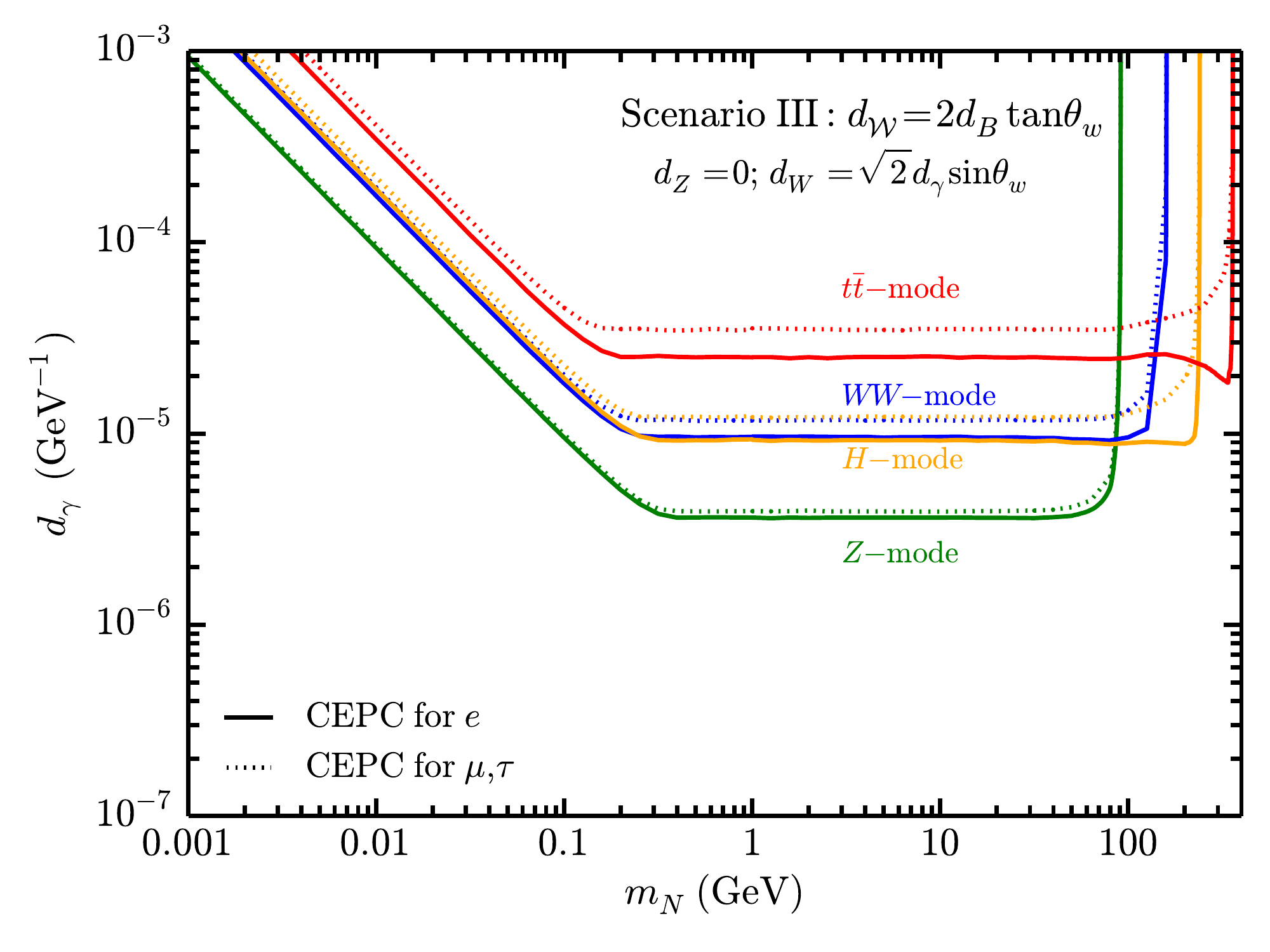}
			\includegraphics[width=0.45\columnwidth]{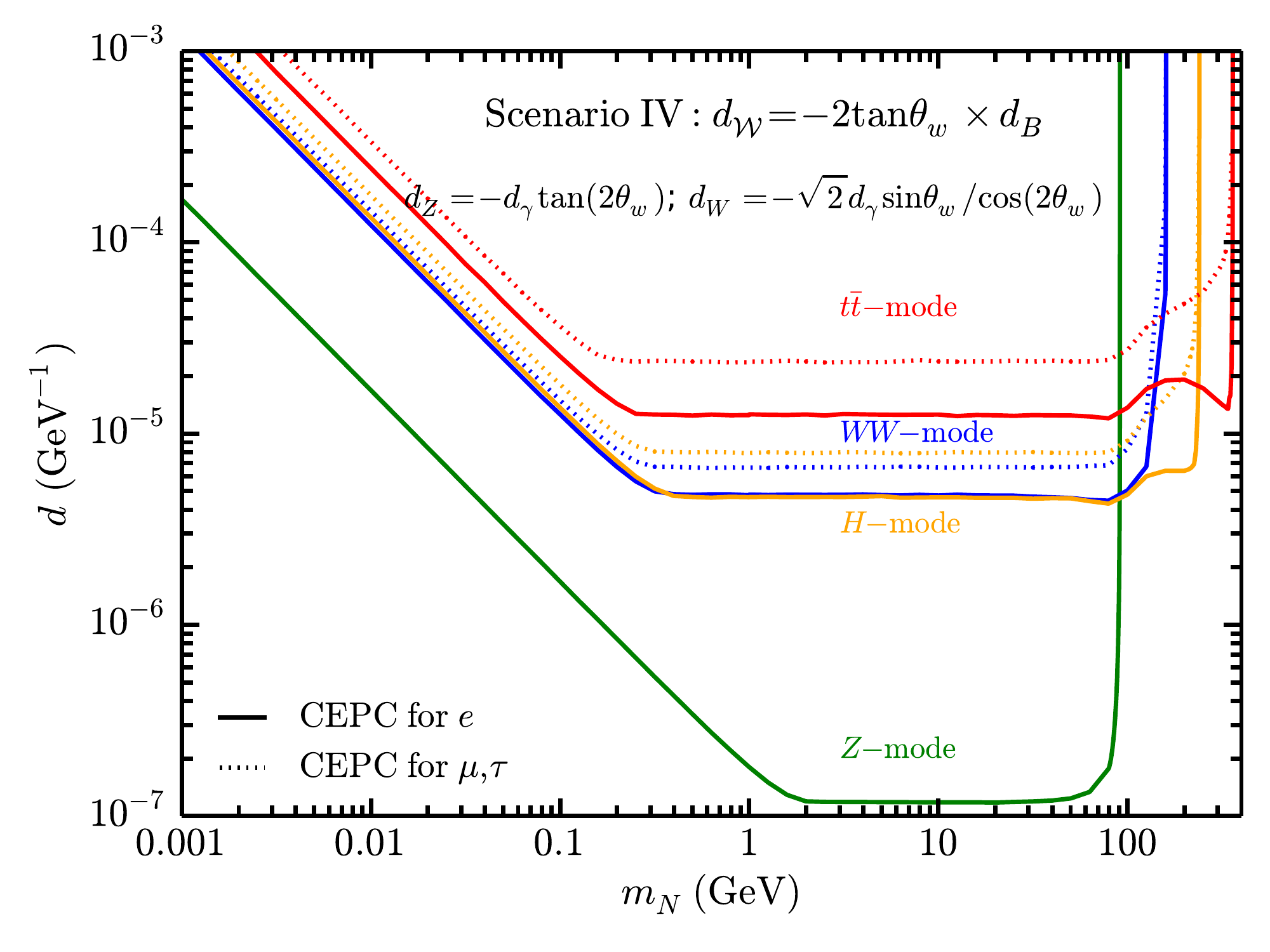}	
			\caption{The expected 95\% C.L. upper limits on the electron-neutrino $d_\gamma^e$ (solid lines), and muon- or tau-neutrino $d_\gamma^{\mu,\tau}$ (dotted lines) dipole portal coupling   to HNLs under four assumptions listed in Table. \ref{tab:scen}  at CEPC in the $Z$-mode with 100 ab$^{-1}$ luminosity (green lines), in the $H$-mode with 20 ab$^{-1}$ luminosity (black lines), in $WW$-mode with 6 ab$^{-1}$ luminosity (blue lines), and in $t\bar t$-mode with 1 ab$^{-1}$ luminosity (red lines), respectively.}
			\label{fig:cepc}
		\end{centering}
	\end{figure}

	The simple criteria $S^2/B=2.71$ is used to probe the 95\% C.L. upper bounds on the neutrino dipole portal 
	couplings $d_\gamma$ at CEPC, which are shown in  Figure \ref{fig:cepc}. Here we consider four scenarios with
	assumptions as $d_\calW=0$, $d_B=0$ and $d_\calW=\pm 2\tan\theta_w d_B$, respectively,
	which are listed in Table. \ref{tab:scen}.
	The limits are calculated based on the total luminosity of $20\ \mathrm{ab}^{-1}$ data in the $H$-mode (orange lines), 
	$6\ \mathrm{ab}^{-1}$  in	the $WW$-mode (blue lines), $100\ \mathrm{ab}^{-1}$  
	in the $Z$-mode (green lines), and $1\ \mathrm{ab}^{-1}$  in the $t\bar t$-mode (red lines).
	There is an additional contribution from $W$-boson, therefore the constraints on $d_\gamma^e$ 
	(plotted with solid lines) are always more stringent than $d_\gamma^{\mu,\tau}$ (plotted with dotted lines) 
	except in scenario I with $d_W=0$ where the sensitivities on all the three lepton flavors are same. 
	In the $Z$-mode at CEPC, the constraints on $d_\gamma$ with different lepton flavor are almost the same, 
	because the additional contribution $t$-channel for electron can be neglected compared to $s$-channel due to the $Z$ resonance when $d_Z\neq 0$. 
	
	One can see that the $Z$-mode has the best sensitivity in all four scenarios for the HNL with small mass. Especially in scenarios I, II, and IV with $d_Z\neq 0$, the upper limits of $d_\gamma^e$ probed by $Z$-mode are stronger than  ones by other three running modes at CEPC beyond one order of magnitude in the mass region of about $(1\sim 50)$ GeV.  What is more, in scenarios II and IV, $Z$-mode can give about two orders of magnitude of improvement over the other three running modes at CEPC in the sensitivity on $d_\gamma^{\mu,\tau}$. This is because  the $Z$-resonance with $\sqrt{s}\simeq M_Z$ can significantly improve the production rate for HNLs at electron colliders.

	%%%%%%%%%%%%%%%%%%%%%%%%%%%%%%%%%%%%%%%%%%%%%%%%%%%%%%%%%%%%%%%
	%%%%%%%%%%%%%%%%%%%%%%%%%%%%%%%%%%%%%%%%%%%%%%%%%%%%%%%%%%%%%%%
	
	In Fig. \ref{fig:dwitha}, we show the 95\% C.L. upper limits on dipole coupling $d_\gamma$ with $m_N=10$ GeV as the function of $a$ in $Z$ mode at CEPC via the monophoton searches, which is plotted with black dashed line. One sees that the curve has similar behavior with  the one from LEP1. The upper bounds on $d_\gamma$ with $a>0$ lie in the range of $(1.5\times10^{-7},3.3\times 10^{-6})$, which is about two order of magnitude lower than LEP1.
	We also present a projection for $d_\gamma$ as the function of $a$ with an imaginary limit on the branching ratio of $10^{-7}$ in the future for the HNL with mass of 10 GeV at 95\% C.L. on the right of Fig.\ref{fig:dwitha} with black dashed line.
	
The graph on the right of Fig. \ref{fig:dwitha100} shows the production rates of the  processes $e^+e^-\to N\nu_e\to  \nu_e\bar\nu_e\gamma$ (red solid line) and $e^+e^-\to N\nu_{\mu,\tau}\to  \nu_{\mu,\tau}\bar\nu_{\mu,\tau}\gamma$ (red dashed line) as the function of $a$ with $m_N=100$ GeV and $d_\gamma=10^{-5}$ at CEPC with $\sqrt{s}=240$ GeV in $H$-mode. 
The corresponding  95\% C.L. upper limits on the dipole portal couplings $d_\gamma^e$ (black solid line) and $d_\gamma^{\mu,\tau}$  (black dashed line) for $m_N=100$ GeV with the luminosity of 20 ab$^{-1}$.
With $|a|\le 10$ and $m_N=100$ GeV, the upper limits on the dipole portal couplings to HNL, $d_\gamma^e$ and $d_\gamma^{\mu,\tau}$, lie in the range of $(2.5\times10^{-6},1.3\times10^{-5})$ and $(9.9\times10^{-6},1.6\times10^{-5})$  from monophoton searches at future CEPC in $H$-mode, respectively.

	\section{DISCUSSION AND CONCLUSION}
	\label{sec:res}
	
\begin{figure}[htbp]
		\begin{centering}
			\includegraphics[width=0.45\columnwidth]{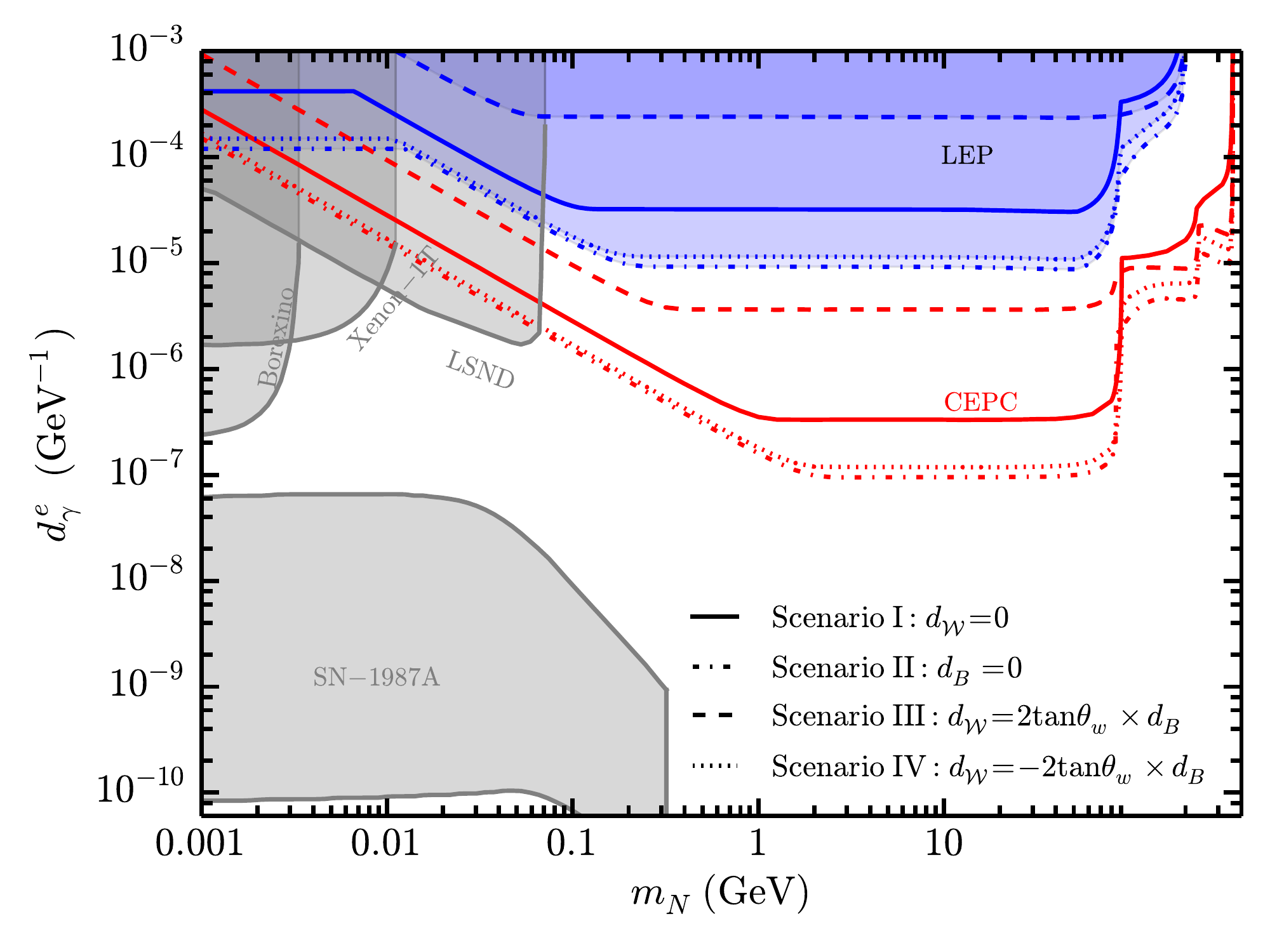}
			\includegraphics[width=0.45\columnwidth]{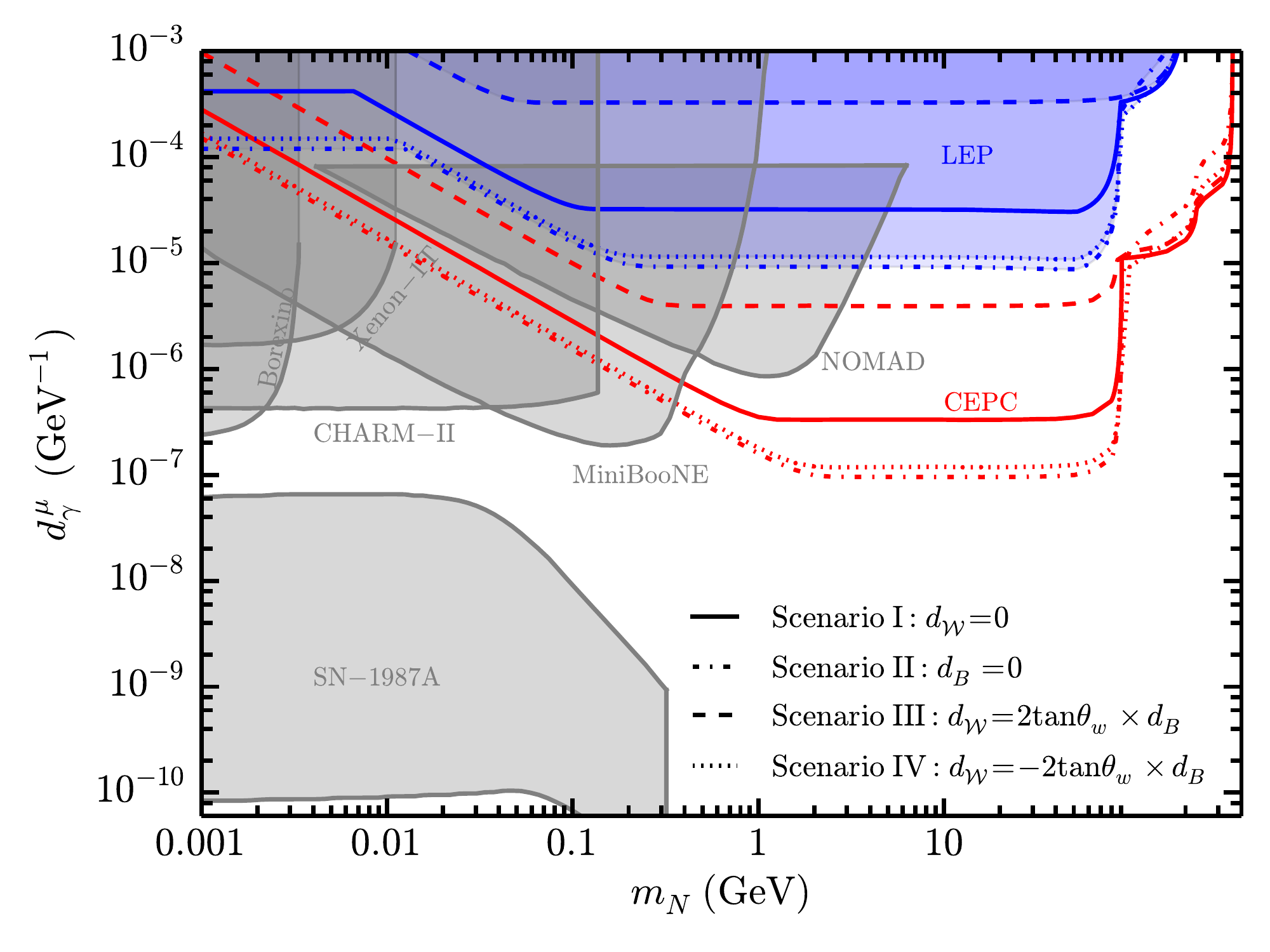}
			\includegraphics[width=0.45\columnwidth]{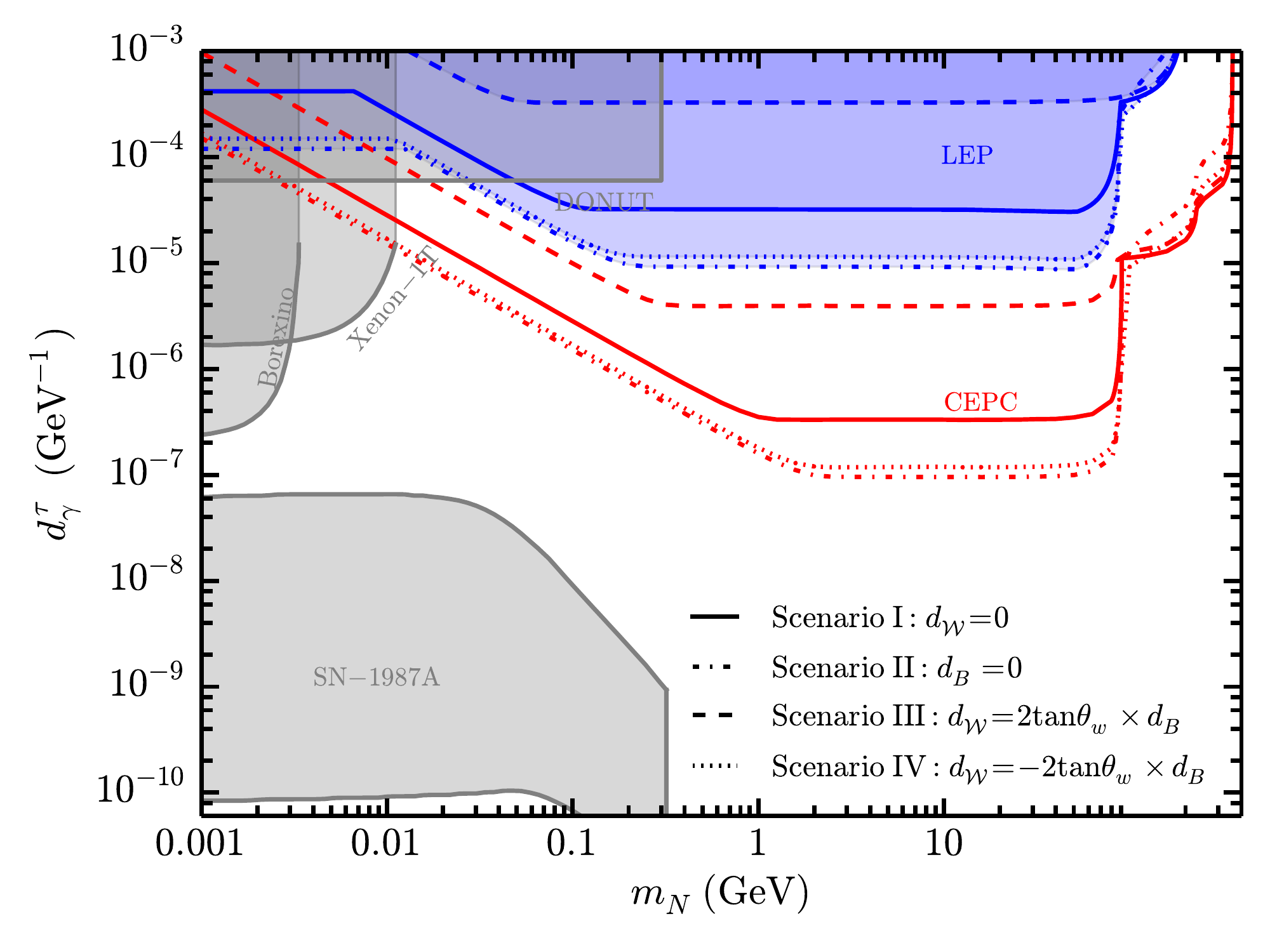}			
			\caption{
				The expected 95\% C.L. exclusion limits on active-sterile neutrino transition magnetic moment $d_\gamma$ in four scenarios listed in Table \ref{tab:scen} at
			LEP (blue shaded regions), which are the combination of the best constraints shown in Fig. \ref{fig:LEP} using the monophoton data by the OPAL Collaboration at LEP1 \cite{OPAL:1994kgw} and by the DELPHI Collaboration at LEP2 \cite{DELPHI:2003dlq}, and the measurement of $Z$ decay \cite{L3:1997exg, ALEPH:2005ab}, and at CEPC (red lines), which are the combination of the best constraints from four running modes at CEPC in Fig. \ref{fig:cepc}, for three lepton flavor respectively.
			The landscape of current leading constraints are also shown with shaded regions, exploiting from
			Borexino \cite{Brdar:2020quo}, Xenon1T \cite{Brdar:2020quo}, LEP \cite{Magill:2018jla}, and SN-1987A \cite{Magill:2018jla}, which are relevant for all three SM neutrinos;  LSND \cite{Magill:2018jla} only for $d_\gamma^e$; CHARM-II \cite{Coloma:2017ppo}, MiniBooNE \cite{Magill:2018jla}, and NOMAD \cite{Magill:2018jla,Gninenko:1998nn} only for $d_\gamma^\mu$; DONUT \cite{DONUT:2001zvi} only for $d_\gamma^\tau$.
			}
			\label{fig:total}
		\end{centering}
	\end{figure}

	The landscape of current constraints on active-sterile neutrino transition magnetic moments $d_\gamma^k$ with $k=e,\mu,\tau$, which are from
	terrestrial experiments such as  Borexino \cite{Brdar:2020quo}, Xenon-1T \cite{Brdar:2020quo}, CHARM-II \cite{Coloma:2017ppo}, MiniBooNE \cite{Magill:2018jla}, LSND \cite{Magill:2018jla}, NOMAD \cite{Gninenko:1998nn,Magill:2018jla},  and DONUT \cite{DONUT:2001zvi}, and  astrophysics supernovae SN 1987A \cite{Magill:2018jla}, are summarized in Fig. \ref{fig:total} with gray shaded regions, respectively. These constraints basically do not dependent on the ratio $a=d_\calW/d_B$, since the typical scattering energies are far less than the electroweak scale. 
	It is noted that the constraints from XENON-1T, Borexino \cite{Brdar:2020quo} and SN 1987A \cite{Magill:2018jla} are flavor-universal. 
	
	The  blue shaded regions in Fig. \ref{fig:total}  present the sensitivities on $d_\gamma$ at LEP in four scenarios listed in Table \ref{tab:scen}, which are the combination of the best constraints shown in Fig. \ref{fig:LEP} using the monophoton data by the OPAL Collaboration at LEP1 \cite{OPAL:1994kgw} and by the DELPHI Collaboration at LEP2 \cite{DELPHI:2003dlq}, and the measurement of $Z$ decay \cite{L3:1997exg, ALEPH:2005ab}.	The combination of the best constraints from four running modes at CEPC in Fig.\ref{fig:cepc} are also shown in Fig. \ref{fig:total} with red lines. It can be found that the constraints from $Z$ invisible decay measured at LEP are already excluded by Xenon-1T.
	The constraints on transition magnetic moments involving
	three SM active neutrinos ($\nu_{e,\mu,\tau}$)  are same from the measurement of $Z$ decay, while the constraints from the monophoton searches at LEP and future CEPC  are  in principle different on  $d_\gamma^e$ and on  $d_\gamma^{\mu,\tau}$ when $a\neq0$ ($d_\calW\neq0$), because there will be additional contributions from $W$-mediator.
	
	For $d_\gamma^e$, beside the  flavor-universal constraints from XENON-1T,
	Borexino \cite{Brdar:2020quo} and SN 1987A \cite{Magill:2018jla}, there are also complementary limits from 
	LSND \cite{Magill:2018jla} with $m_N\lesssim0.07$ GeV. For heavier $N$ only coupling with $\nu_e$, LEP can explore the  previously unconstrained parameter region, and will be greatly improved by CEPC.
	In scenario III, with $d_Z=0$, the contribution from $Z$ boson vanishes, leading to 
	weakest limits on $d_\gamma^e$ than other three scenarios  which can be down to about $3.3\times 10^{-4}$ at LEP and $3.6\times 10^{-6}$ at CEPC.
	In  scenario II, with $d_B=0$, CEPC can probe the limit on $d_\gamma^e$ down to about $9.5\times 10^{-8}$, which is more than two orders of magnitude stronger than LEP, with the limit down to about $1.3\times 10^{-5}$ from the measurement of ${\rm Br}(Z\to\gamma+{\rm invisible})$. 
	
	For $d_\gamma^\mu$, there are terrestrial constraints from CHARM-II \cite{Coloma:2017ppo}, MiniBooNE, and NOMAD \cite{Magill:2018jla}. In scenario III, the  best limit on $d_\gamma^\mu$ at LEP is  from the monophoton searches by the DELPHI Collaboration at LEP2 \cite{DELPHI:2003dlq} in the plotted region, which is weaker than the one on $d_\gamma^e$ due to the absence of $W$-exchanging channel,  while  still can touch the unexplored parameter region when $m_N\gtrsim 1$ GeV.
	In other three scenarios with $d_Z\neq 0$, the limits on $d_\gamma^\mu$ with $m_N\lesssim 90$ GeV shown in Fig. \ref{fig:total} are from the $Z$ decay measurements at LEP, thus they are same with $d_\gamma^e$, which are  ahead of the 
	current limits from  NOMAD \cite{Magill:2018jla} with $m_N$ larger than 5.0 GeV, 3.6 GeV, and 3.9 GeV in scenario I, II, and IV, respectively.
	The expected limits from the monophoton searches at CEPC are complementary to the current limits when $m_N \gtrsim 3.5$ GeV  in scenario III and $m_N \gtrsim 0.4$ GeV  in other three scenarios.
	
	On $d_\gamma^\tau$, there is an upper 90\% C.L. limit given by DONUT \cite{DONUT:2001zvi}
	of 5.8 $\times 10^{-5}\ {\rm GeV}^{-1}$ for $m_N<0.3$ GeV. The constraints on $d_\gamma^\tau$ from LEP and CEPC are same with $d_\gamma^\mu$. The monophoton searches at LEP2 can provide complementary constraints   to DONUT on $d_\gamma^\tau$ when $m_N>0.3$ GeV in scenario III with $d_Z=0$.
	In other scenarios with $d_Z\neq 0$, the measurement of $Z\to\gamma+ {\rm invisible}$ at LEP can provide leading sensitivity on $d_\gamma^\tau$ with $m_N \gtrsim 0.05\ (0.03) $ GeV in  scenario I (II/IV), respectively.
	The monophoton searches at CEPC can fill a huge unconstrained  void of the $d_\gamma^\tau-m_N$ parameter space when $m_N \gtrsim 0.2$ GeV.

	%%%%%%%%%%%%%%%%%%%%%%%%%%%%%%%%%%%%%%%%%%%%%%%%%%%%%%%%%%%%%%%
	%%%%%%%%%%%%%%%%%%%%%%%%%%%%%%%%%%%%%%%%%%%%%%%%%%%%%%%%%%%%%%%
	\acknowledgments
	This work was supported in part by the National Natural Science Foundation of China (Grant No.12205153, No. 11805001) and 
	the 2021 Jiangsu Shuangchuang (Mass Innovation and Entrepreneurship) Talent Program (JSSCBS20210213).

\bibliographystyle{JHEP}
\bibliography{submit.bib}
\end{document}